%% file: paper.tex
\documentclass[11pt,a4paper,fleqn]{article}
\pdfoutput=1   
\usepackage[utf8]{inputenc}
\usepackage{lmodern}
\usepackage{alpha}
\hypersetup{%
   pdftitle    = {openQ\*D code: a versatile tool for QCD+QED simulations},
   pdfauthor   = {I. Campos, P. Fritzsch, M. Hansen, M. Krstic Marinkovic, A. Patella, A. Ramos, N. Tantalo},
   pdfkeywords = {Lattice QCD and QED, high performance computing, C* boundary conditions}
}%
\usepackage{defs}
\begin{document}

\input{title.tex}
\tableofcontents
\pagebreak

\section{Introduction}\label{sec:intro}
\input{10-introduction.tex}

\section{Theoretical background}\label{sec:tech}
\input{20-thintro.tex}

\subsection{\Cstar boundary conditions}\label{sec:cstar}
\input{22-cstar.tex}

\subsection{Gauge actions}\label{subsec:gauge}
\input{24-definitions.tex}

\subsection{Dirac operator}\label{subsec:dirac}
\input{25-dirac.tex}

\section{Simulating QCD+QED with \oQxD}\label{sec:guide}
\input{30-userguide.tex}

\clearpage
\section{Performance and testing} \label{sec:perftests}

\subsection{Code performance on parallel machines}\label{sec:perform}
\input{40-performance.tex}

\subsection{Low-level tests}\label{sec:tests}
\input{42-tests.tex}

\subsection{Conservation of the Hamiltonian with Fourier Acceleration}\label{sec:dH}
\input{44-hamiltonian.tex}

\subsection{Performance of locally deflated solver in QCD+QED}\label{sec:dfltests}
\input{46-deflation.tex}

\subsection{Key observables for HMC simulations of \QCDQED}\label{sec:obs}
\input{48-observables.tex}

\section{Summary and outlook}\label{sec:summary}
\input{90-summary.tex}

\begin{acknowledgement}%
\input{acknow.tex}
\end{acknowledgement}
\clearpage

\appendix
\section{Implementation of the RHMC}\label{sec:rhmc}
\input{98-rhmc.tex}

\section{Laplacian for the Fourier Accelerated Molecular Dynamics}\label{sec:laplacian}
\input{99-fourier.tex}

\clearpage
\section{Sample input file}\label{app:pedro01.in}
\begin{lstlisting}[basicstyle=\scriptsize\ttfamily]
[Run name]
name       pedro01

[Directories]
log_dir    ./log      # absolute path, or relative path
dat_dir    ./dat      # to the working directory of iso1
cnfg_dir   ./cnfg

[MD trajectories]
nth         100       # multiple of dtr_cnfg
ntr         800       # multiple of dtr_cnfg
dtr_log     5
dtr_ms      10        # multiple of dtr_log
dtr_cnfg    50        # multiple of dtr_ms

[Random number generator]
level      0          # this should not be changed
seed       19521      # this can be any positive integer

[Boundary conditions]
type       periodic   # or SF, open, open-SF
cstar      3          # or 0, 1, 2

[SU(3) action]
beta       5.3
c0         1.0      # 1=Wilson, 5/3=Lüscher-Weisz, 3.648=Iwasaki

[U(1) action]
type       compact  # only option currently available
alpha      0.05     # bare fine-structure constant
invqel     6.0      # see "Dirac operators parameters" 
c0         1.0      # Wilson action

[Quark action]
nfl        2

[Flavour 0]          # Down quark
qhat       -2        # qhat must be integer
                     # el. charge = qhat/invqel = -2/6 = -1/3
kappa      0.136377  # hopping parameter
su3csw     1.909520  # u1csw=su3csw=0 => no O(a) improv.
u1csw      1.0       # u1csw=su3csw=1 => tree-level O(a) improv.

[Flavour 1]          # Up quark
qhat       4         # el. charge = qhat/invqel = 4/6 = 2/3
kappa      0.137312
su3csw     1.909520
u1csw      1.0

[Rational 0]
power		 -1 4
degree		 10
range		 1.98000000e-03 7.62000000e+00
mu		 0.00000000e+00
delta		 5.9691841082503071e-05
A		 2.04978213590663732591e-01
nu[0]		 1.22647978559899293316e+01
mu[0]		 8.40737261524814627478e+00
nu[1]		 3.58475041480018230544e+00
mu[1]		 2.79734086059047637463e+00
nu[2]		 1.37354614313178191587e+00
mu[2]		 1.08904089757432842589e+00
nu[3]		 5.45534380719472244969e-01
mu[3]		 4.33597479995857904012e-01
nu[4]		 2.17882243098165673256e-01
mu[4]		 1.73241240349149644429e-01
nu[5]		 8.70901176278380123597e-02
mu[5]		 6.92465791863651897176e-02
nu[6]		 3.47963276911667715452e-02
mu[6]		 2.76565520583723425951e-02
nu[7]		 1.38540251643490298916e-02
mu[7]		 1.09844143754786495448e-02
nu[8]		 5.39355078694815723989e-03
mu[8]		 4.20882858056409459024e-03
nu[9]		 1.79456777883689466702e-03
mu[9]		 1.23015480378516474207e-03
x[0]		 3.92039999999999976796e-06
x[1]		 4.81832134522929173714e-06
x[2]		 8.31937464355889515232e-06
x[3]		 1.75765908325188680071e-05
x[4]		 4.09284748868167719878e-05
x[5]		 9.94161232736990132294e-05
x[6]		 2.45725878382011789398e-04
x[7]		 6.11688882726808629518e-04
x[8]		 1.52696110343796594838e-03
x[9]		 3.81619175022324466640e-03
x[10]		 9.54119167823534904127e-03
x[11]		 2.38582015157776800018e-02
x[12]		 5.96499569883204849852e-02
x[13]		 1.49077585046584693007e-01
x[14]		 3.72142898437580804671e-01
x[15]		 9.26380547538878773572e-01
x[16]		 2.28972591430973704263e+00
x[17]		 5.56179223363444652506e+00
x[18]		 1.29510708833971612819e+01
x[19]		 2.73621135618227562247e+01
x[20]		 4.72437717308978761821e+01
x[21]		 5.80643999999999991246e+01

[HMC parameters]
actions    0 1 2 3  # List of action IDs, see below
npf        2        # Number of pseudofermions to be allocated
nlv        2        # Number of levels of integrator for MD eqs
tau        2.0      # MD trajectory length
facc       1        # Fourier acceleration for U(1) MD
                    # (0=not active, 1=active)

[Level 0]           # Innermost level
integrator OMF4     # Omelyan-Mryglod-Folk 4th order
nstep      2        # Number of times the elementary integrator
                    # is applied at this level
forces     0 1      # List of force IDs to be integrated at
                    # this level, see below

[Level 1]           # Outermost level
integrator OMF4
nstep      1
forces     2 3

[Action 0]          # No adjustable parameters here!
action     ACG_SU3

[Force 0]
force      FRG_SU3

[Action 1]
action     ACG_U1

[Force 1]
force      FRG_U1

[Action 2]
action  ACF_RAT_SDET # Rational approximation effective action
ipf     0            # Pseudofermion ID (a number from 0 to 1)
ifl     0            # Flavour ID (down quark)
irat    0 0 9        # Use the rational approximation with ID=0
                     # Include all rat. appr. factors, 0 -> 9,
                     # i.e. no frequency splitting
isp     0            # Solver ID, used to generate the p.f. at
                     # the beginning of the MD and to calculate
                     # the Hamiltonian at the end of the MD

[Force 2]
force   FRF_RAT_SDET
isp     1            # Solver ID, used to calculate the force

[Action 3]
action  ACF_RAT_SDET
ipf     1            # Different pseudofermion ID
ifl     1            # Different flavour ID (up quark)
irat    0 0 9
isp     0

[Force 3]
force   FRF_RAT_SDET
isp     1

[Solver 0]
solver     MSCG      # or CGNE, SAP_GCR, DFL_SAP_GCR
nmx        2048      # Maximum number of iterations
res        1.0e-11   # Residue

[Solver 1]
solver     MSCG
nmx        2048
res        1.0e-8

[Wilson flow]
integrator RK3       # EULER: Euler, RK2: 2nd order Runge-Kutta
                     # RK3: 3rd order Runge-Kutta
eps        2.0e-2    # Integration step size
nstep      700       # Total number of integration steps
dnms       5         # Number of steps between measurements
\end{lstlisting}

\clearpage
\small
\addcontentsline{toc}{section}{References}
\bibliographystyle{JHEP}
\bibliography{lattice}

\end{document}

%% file: title.tex
\preprintno{%
CERN-TH-2019-136\\
\vfill
}

\title{%
\texttt{openQ*D} code: a versatile tool for QCD+QED simulations
}

\collaboration{\RClogo~collaboration}

\author[ifca]{Isabel Campos} 
\author[cern]{Patrick Fritzsch}
\author[TVinfn]{Martin Hansen}
\author[trin]{Marina Krstic Marinkovic}
\author[hu]{Agostino Patella}
\author[trin]{Alberto Ramos}
\author[TVrome,TVinfn]{Nazario Tantalo}

\address[ifca]{Instituto de F\'isica de Cantabria \& IFCA-CSIC, Avda. de Los Castros s/n, 39005 Santander, Spain}
\address[cern]{Theoretical Physics Department, CERN, CH-1211 Geneva 23, Switzerland}
\address[TVinfn]{INFN, Sezione di Tor Vergata, Via della Ricerca Scientifica 1, 00133 Rome, Italy}
\address[trin]{School of Mathematics, Trinity College Dublin, Dublin 2, Ireland}
\address[hu]{Humboldt Universit\"at zu Berlin, Institut f\"ur Physik \& IRIS Adlershof, \\Zum Grossen Windkanal 6, 12489 Berlin, Germany}
\address[TVrome]{Universit\`a di Roma Tor Vergata, Dipartimento di Fisica, \\Via della Ricerca Scientifica 1, 00133 Rome, Italy}

\begin{abstract}
We present the open--source package \texttt{openQ*D-1.0}~\cite{openQxD-csic},
which has been primarily, but not uniquely, designed to perform lattice
simulations of QCD+QED and QCD, with and without \Cstar boundary conditions,
and $O(a)$ improved Wilson fermions. The use of \Cstar boundary conditions in
the spatial direction allows for a local and gauge--invariant formulation of
QCD+QED in finite volume, and provides a theoretically clean setup to calculate
isospin--breaking and radiative corrections to hadronic observables from first
principles. The \texttt{openQ*D} code is based on
\texttt{openQCD-1.6}~\cite{openQCD} and \texttt{NSPT-1.4}~\cite{NSPT}. In
particular it inherits from \texttt{openQCD-1.6} several core features, e.g.
the highly optimized Dirac operator, the locally deflated solver, the frequency
splitting for the RHMC, or the 4th order OMF integrator. \\
\end{abstract}

\begin{keyword}
Lattice QCD and QED \sep Gauge Invariance \sep High Performance Computing  %
\PACS{%
11.15.-q\sep 
11.15.Ha\sep 
12.20.-m\sep 
12.38.Gc\sep 
12.38.-t\sep 
02.70.-c\sep 
02.70.Uu     
}
\end{keyword}

\maketitle
\makeatletter
\g@addto@macro\bfseries{\boldmath}
\makeatother

%% file: 10-introduction.tex
QED radiative corrections to hadronic observables are generally rather small but
they become phenomenologically relevant when the target precision is at the
percent level. For example, the leptonic and semileptonic decay rates of light
pseudoscalar mesons are measured with a very high accuracy and, on the
theoretical side, have been calculated with the required non--perturbative
accuracy by many lattice collaborations. Most of these calculations have been
performed by simulations of lattice QCD without taking into account QED
radiative corrections. A recent review~\cite{Aoki:2019cca} of the results
obtained by the different lattice groups shows that leptonic and semileptonic
decay rates of $\pi$ and $K$ mesons are presently known at the sub--percent
level of accuracy. At the same time, QED radiative corrections to these
quantities are estimated to be of the order of a few percent, by means of
chiral perturbation theory~\cite{Cirigliano:2011ny}. These estimates have
recently been confirmed in the case of the leptonic decay rates of $\pi$ and $K$
by a first--principle lattice calculation of the QED radiative corrections at
$O(\alpha)$ in refs.~\cite{Giusti:2017dwk,DiCarlo:2019thl}.

Other remarkable examples of observables for which QED radiative corrections are
phenomenologically relevant are the so--called lepton flavour universality
ratios. For example $R(D^{(*)})$ is defined as the branching ratio for $B
\mapsto D^{(*)} \ell \bar \nu_\ell$ with $\ell=e,\mu$ divided by the branching
ratio for $B \mapsto D^{(*)} \tau \bar \nu_\tau$. Most of the hadronic
uncertainties cancel in these ratios that are built in such a way that they are trivial
in the Standard Model, in the limit in which the two leptons have the same mass.
Presently, a combined analysis~\cite{HFLAV} of the $R(D)$ and $R(D^*)$ ratios
shows a deviation of the experimental measurements from the theoretical
predictions of the order of $3$ standard deviations. On the other hand, QED
radiative corrections are different for the two leptons because of the different
masses and an improved theoretical treatment of these effects (see for example
refs.~\cite{deBoer:2018ipi,Cali:2019nwp} for a discussion of this point) can
possibly enhance or reconcile the observed discrepancy between the experimental
measurements and the theoretical expectations.

QED radiative corrections to hadronic observables can be computed from first
principles by performing lattice simulations of QCD coupled to QED, treating the
photon field on an equal footing as the gluon field. This approach, pioneered in
refs.~\cite{Borsanyi:2014jba, Horsley:2015eaa, Horsley:2015vla}, is highly
non--trivial from both the numerical and theoretical point of view, because of
the peculiarities of QED. Numerically, lattice calculations are unavoidably
affected by statistical and systematic uncertainties and it can be challenging
to resolve QED radiative corrections from the leading QCD contributions within
the errors of a simulation. Theoretically, a big issue arises because lattice
calculations have necessarily to be done on a finite volume. QED is a
long--range interaction and, consequently, finite--volume effects are the key
issue in presence of electromagnetic interactions.

In fact, as a consequence of Gauss' law, it is impossible to have a net
electric charge on a periodic torus. Because of this strong theoretical
constraint, it is particularly challenging to calculate from first principles
physical observables associated with electrically charged external states,
such as the phenomenologically relevant quantities discussed above. Several
approaches have been proposed over the years to cope with this problem, see
ref.~\cite{Patella:2017fgk} for a recent review. The most popular approaches
to the problem of charged particles on the torus solve the Gauss' law
constraint by introducing non--local terms in the finite--volume action of the
theory.%
\footnote{%
    A different approach is based on the idea that one can write QCD+QED
    observables at first order in $\alpha$ as QCD observables with analytic
    (possibly infinite--volume) QED kernels,
    e.g.~\cite{Bernecker:2011gh,Blum:2017cer,Feng:2018qpx}.
}%
The effects induced by the non--locality of the action are expected to
disappear once the infinite--volume limit is properly taken and, as far as
$O(\alpha)$ QED radiative corrections are concerned, it is generally possible
to show that this is indeed the case. 

On the one hand, the non--local formulations of the theory are particularly
appealing because of their formal simplicity. On the other hand, it has been
shown in ref.~\cite{Lucini:2015hfa} that it is possible to probe electrically
charged states on a finite volume by starting from a local formulation of the
theory and, remarkably, in a fully gauge--invariant way. This is possible by
using C--parity (or \Cstar) boundary conditions for all the fields and by using a certain
class of interpolating operators originally introduced by Dirac in a seminal
work~\cite{Dirac:1955uv} on the canonical quantization of QED. 

The formulation of ref.~\cite{Lucini:2015hfa} has also been studied
numerically. The results for the meson masses extracted in a fully
gauge--invariant way from lattice simulations of QCD$+$QED with \Cstar
boundary conditions obtained in ref.~\cite{Hansen:2018zre} provide a convincing
numerical evidence that, beside being an attractive theoretical formulation,
the proposal of ref.~\cite{Lucini:2015hfa} is also a valid numerical
alternative for the calculation of QED radiative corrections on the lattice.
This motivated the present work.

In this paper we present the open--source package \oQxD, which can be used to
simulate QCD$+$QED, QCD, the pure SU(3) and U(1) gauge theories.%
\footnote{%
        The code allows also for (inefficient) simulations of QED in isolation,
        even though a main program for this purpose is not provided in the 1.0
        version.
}%
The code allows to choose a
wide variety of temporal and spatial boundary conditions. In
particular, it allows to perform dynamical simulations of QCD$+$QED with
\Cstar but also with periodic boundary conditions along the spatial directions.
Simulations of QCD with \Cstar boundary conditions can be a valuable starting
point for the application of the RM123 method~\cite{deDivitiis:2013xla}, in
which observables are calculated order--by--order in the electromagnetic
coupling. A fully tested and stable relase of \oQxD can be
downloaded from~\cite{openQxD-csic}.

The \oQxD package is based on the \oQCD~\cite{openQCD} package from which it
inherits the core features, most notably the implementation of the Dirac
operator, of the solvers and the possibility of simulating open and
Schr\"odinger functional boundary conditions in the time direction. One of the
inherited solvers implements the inexact deflation algorithm of
ref.~\cite{Luscher:2007es}. An added value of the \oQxD package is the
possibility of using more deflation subspaces in a single simulation. This is
particularly important in the case of QCD$+$QED simulations because different
deflation subspaces have to be generated for quarks having different electric
charges.

Another important feature present in the \oQxD package is the possibility to use
Fourier Acceleration~\cite{Batrouni:1985jn,Duane:1988vr} for the
molecular dynamics evolution of the U(1) field. The used implementation of the Fast
Fourier Transform (FFT) is an adaptation of the
corresponding module in the \texttt{NSPT}~\cite{DallaBrida:2017tru,NSPT}
package.

\begin{figure}[t]
 \centering
 \includegraphics[width=0.9\textwidth,clip]{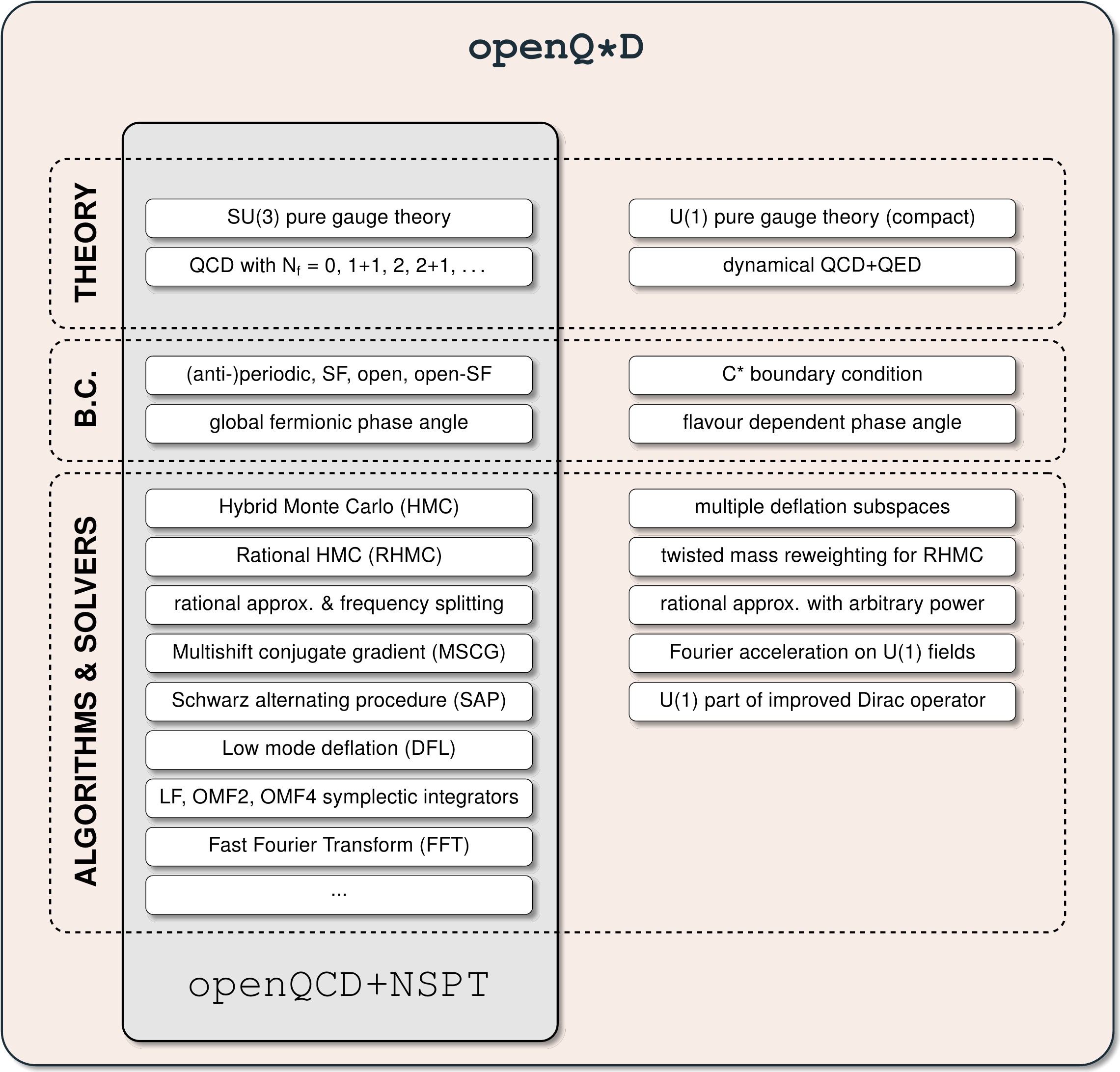}
 \caption{%
         Summary of salient features of \oQxD. Some features inherited from
         \oQCD and \texttt{NSPT} are highlighted.
         }\label{fig:overview}
\end{figure}

The remaining of this paper is organised as follows. In section~\ref{sec:tech}
we give an overview of the theoretical background needed to understand the
actions simulated by \oQxD, and we describe some peculiar aspects of the
simulation algorithm. In particular, the specific implementation of \Cstar
boundary conditions and of the Fourier Acceleration for the U(1) field are
discussed. In section~\ref{sec:guide} we provide instructions on how to compile
the code, construct a sample input file, and run the program that generates
QCD$+$QED configurations. Section~\ref{sec:perftests} is a collection of tests and
performance studies. In particular, we present scalability tests, and studies of
the performance of solvers for the Dirac equation for electrically charged
fields. We also illustrate the outcome of some sample runs performed for testing
purposes. In figure~\ref{fig:overview}, we provide a schematic view of the
\oQxD functionalities.
%

%% file: 20-thintro.tex
An overview of the main algorithmic choices made in the code will be given in
this section. The fundamental fields are the SU(3) link variable $U_{\mu}(x)$
and the real photon field $A_{\mu}(x)$. Since only the
compact formulation of QED is implemented at present, all observables are written in terms of the U(1)
link variable
\begin{align}
    z_{\mu}(x) &= \exp\{i A_{\mu}(x)\}   \;,
\end{align}
which implies that the real photon field can be restricted to $-\pi \leq
A_{\mu}(x) \leq \pi$ with no loss of generality. Various boundary conditions
can be chosen for the gauge fields: periodic, open~\cite{Luescher2013},
Schr\"odinger Functional (SF)~\cite{Luscher:1992an,Sint:1993un} and open--SF
boundary conditions~\cite{Luscher:2014kea} in the Euclidean time direction
$\mu=0$, periodic and \Cstar boundary
conditions~\cite{Kronfeld:1990qu,Kronfeld:1992ae,Wiese:1991ku,Polley:1993bn} in
the spatial directions. The implementation of \Cstar boundary conditions is
discussed in section~\ref{sec:cstar}.

After integrating out the fermion fields in a usual way, the target distribution
of QCD+QED if no \Cstar boundary conditions are used is
\begin{align}\label{eq:firstdistr-det}
 \rho_\text{tar}(U,A) &\propto e^{-\SgSU(U)-\SgU(A)} \prod_{f} \det D_f \;,
\end{align}
where the gauge actions $\SgSU(U)$ and $\SgU(A)$ are briefly discussed in
section~\ref{subsec:gauge}, the product runs over the simulated fermion
flavours indicized by $f$, and the Dirac operator $D$ is introduced in
section~\ref{subsec:dirac}. If \Cstar boundary conditions are used, the
determinant is replaced by a Pfaffian, i.e.
\begin{align}\label{eq:firstdistr-pf}
    \rho_\text{tar}(U,A) &\propto e^{-\SgSU(U)-\SgU(A)} \prod_{f} \text{pf} \, (CTD_f) \;,
\end{align}
where $C$ is the charge conjugation matrix and $T$ is a field--independent matrix
satisfying $T^2=1$, whose detailed definition can be found in
section~\ref{sec:cstar}. While in the continuum limit the determinant and the
Pfaffian are positive, this is not the case with Wilson fermions. The absolute
value is considered in both cases, which amounts to replacing
\begin{align}
   \det D_f &\rightarrow \left| \det D_f \right| \;, &
   \text{pf} \, (CTD_f) &\rightarrow \left| \text{pf} \, (CTD_f) \right| = \left| \det D_f \right|^{1/2}  \;.
\end{align}
The sign should be separately calculated and included in the evaluation of
observables as a reweighting factor~\cite{Montvay:2001aj,Ali:2018dnd}. It is
important to stress that this is a mild sign problem~\cite{Lucini:2015hfa},
which becomes irrelevant sufficiently close to the continuum limit, and which
is also present in standard QCD simulations for the strange quark. The
presented strategy is in line with state--of--the--art QCD and QCD+QED
simulations, in which the sign of the determinant is simply ignored.  Future
work will be planned to investigate the importance of the sign especially at
lighter quark masses.

After introducing the standard even--odd preconditioned operator
$\Dhat$~\cite{DeGrand:1988vx}, one rewrites the quark part of the distribution
as
\begin{align}\label{eq:physdistr}
    \prod_{f} \left| \det D_f \right|^{2\alpha_f} &=  
    \prod_{f} \det (D^{\dagger}_f D_f)^{\alpha_f}  = 
    e^{-\Ssdet(U,A)} \prod_{f} \det (\Dhat^{\dagger}_f \Dhat_f)^{\alpha_f} \;,
\end{align}
where $\alpha_f$ is either $1/2$ or $1/4$. The definitions of $\Dhat_f$ and
$\Ssdet$ can be found in section~\ref{subsec:dirac}. Instead of this
target distribution, the \oQxD code simulates a slightly different
distribution
\begin{align}\label{eq:simdistr}
   \rho_\mathrm{sim}(U,A) &\propto
   e^{-\SgSU(U)-\SgU(A)} e^{-\Ssdet(U,A)} \prod_{f} \det \Rmu_f^{-1} \;.
\end{align}
written in terms of a rational approximation $\Rmu_f$~\cite{Kennedy:1998cu}
\begin{align}\label{eq:R}
   \Rmu_f &\simeq ( \Dhat_f^\dag \Dhat_f + {\mu}_f^2 )^{-\alpha_f} \;,
\end{align}
where ${\mu}_f$ is a tunable parameter introduced to suppress configurations
with exceptionally small eigenvalues of $\Dhat_f^\dag \Dhat_f$
(\textit{twisted--mass reweighting}~\cite{Luscher:2008tw,Luscher:2012av}). If ${\mu}_f$ is
small enough and the rational approximation is accurate enough, the simulated
distribution $\rho_\mathrm{sim}(U,A)$ is very close to the target one
$\rho_\mathrm{tar}(U,A)$. The difference is corrected by means of reweighting
factors $W_f$
\begin{align}
   \frac{\rho_\mathrm{tar}(U,A)}{\rho_\mathrm{sim}(U,A)} &\propto \prod_f W_f  \;, & \label{eq:W}
   W_f &= \det\left[ ( \Dhat_f^\dag \Dhat_f)^{\alpha_f} \Rmu_f \right]  \;,
\end{align}
which have to be separately calculated and included in the expectation values of
observables as follows
\begin{align}
   \langle O \rangle_\mathrm{tar} &=
   \frac{ \langle O \prod_f W_f \rangle }{ \langle \prod_f W_f \rangle }  \;.
\end{align}
The detailed discussion of the supported reweighting factors can be found in
appendix~\ref{sec:rhmc}.
The rational function $\Rmu_f$ can be decomposed in a product of positive
factors $\Rmu_{f,\ell}$ (\textit{frequency splitting}~\cite{Luscher:2012av}).
More details on frequency splitting are provided in section~\ref{subsec:freq}.
The determinant of the rational functions is finally represented by means of a
pseudofermion quadratic action as in
\begin{align}
        \det \Rmu_f^{-1} &= \prod_\ell \det \Rmu_{f,\ell}^{-1} =
   \int [d\Phi] \, e^{- \sum_\ell ( \Phi_{f,\ell}, \Rmu_{f,\ell} \Phi_{f,\ell} ) } \;.
\end{align}
The distribution is generated by means of a Hybrid Monte Carlo (HMC) algorithm
with Fourier acceleration for the U(1) field. The molecular dynamics (MD)
Hamiltonian is given by
\begin{gather}\label{eq:Hamiltonian}
   H = \frac{1}{2} ( \pi , \Delta^{-1} \pi )_\text{U(1)} + \frac{1}{2} ( \Pi , \Pi )_\text{SU(3)} + S(U,A,\Phi) \ ,
\end{gather}
where $\Pi_\mu(x)$ and $\pi_\mu(x)$ denote the momentum fields associated to the
SU(3) and U(1) fields, the operator $(-\Delta)$ is a discretization of the
Laplace operator, and the action is given by
\begin{align}\label{eq:action}
   S(U,A,\Phi) &= \SgSU(U) + \SgU(A) + \Ssdet(U,A) + \sum_{f,\ell} ( \Phi_{f,\ell}, \Rmu_{f,\ell} \Phi_{f,\ell} ) \;.
\end{align}
Details on the implementation of the Fourier acceleration are  presented in
appendix~\ref{sec:laplacian}. The HMC consists of three steps.
\begin{enumerate}\setlength{\itemsep}{0em}
   \item The momentum and pseudofermion fields are randomly generated with
   probability distribution given by $e^{-H}$;
   \item The gauge fields are evolved with a discretized version of the MD equations, i.e.
   \begin{alignat}{2}
      &
      \partial_t A_\mu(x) = \Delta^{-1} \pi_\mu(x)
      &\hspace{1cm}&
      \partial_t \pi_\mu(x) = - \partial_{A_\mu(x)} S(U,A,\Phi)
      \ ,
      \nonumber
      \\
      &
      \partial_t U_\mu(x) = \Pi_\mu(x) U_\mu(x)
      &&
      \partial_t \Pi_\mu(x) = - \partial_{U_\mu(x)} S(U,A,\Phi)
      \ ,
   \end{alignat}
   where $\partial_{U_\mu(x)}$ is the left Lie derivative with respect to
   $U_\mu(x)$ while $\partial_{A_\mu(x)}$ is the elementary derivative with
   respect to $A_\mu(x)$. In practice multiple time--scale~\cite{Sexton:1992nu}
   symplectic integrators are used to solve the MD equation: leapfrog, 2nd and
   4th order Omelyan--Mryglod--Folk integrators~\cite{OMELYAN2003272} are
   available (LF, OMF2, OMF4).
   \item The evolved gauge configutation is accepted or rejected with a standard
   Metropolis test with probability distribution given by $e^{-H}$.
\end{enumerate}

%% file: 22-cstar.tex
\begin{figure}[t]
\centering
\begin{tikzpicture}[scale=.60]

   \draw[-latex] (-1.5,-1) -- ++(3,0) node[anchor=north] {1};
   \draw[-latex] (-1,-1.5) -- ++(0,3) node[anchor=east,text width=12mm,align=right] {\Cstar dir.\\$k \neq 1$};
   
   \foreach \x in {0,1,...,5} {
      \foreach \y in {0,1,...,5} {
         \draw[fill=black] (\x,\y) circle (.1);
         \draw[thick] (\x,\y) -- ++(1,0);
         \draw[thick] (\x,\y) -- ++(0,1);
      }
   }
   \foreach \x in {6,7,...,11} {
      \foreach \y in {0,1,...,5} {
         \draw[black!40,fill=black!40] (\x,\y) circle (.1);
         \draw[thick,black!40] (\x,\y) -- ++(1,0);
         \draw[thick,black!40] (\x,\y) -- ++(0,1);
      }
   }
   
   \foreach \x in {0,1,...,5} {
      \draw[red,thick] (\x,6) circle (.1);
   }
   \draw[red,thick] (-.3,5.7) rectangle (5.3,6.3);
   \draw[red,thick] (5.7,-.3) rectangle (11.3,.3);
   
   \foreach \x in {6,7,...,11} {
      \draw[green!70!black,thick] (\x,6) circle (.1);
   }
   \draw[green!70!black,thick] (5.7,5.7) rectangle (11.3,6.3);
   \draw[green!70!black,thick] (-.3,-.3) rectangle (5.3,.3);
   
   \foreach \y in {0,1,...,5} {
      \draw[blue,thick] (12,\y) circle (.1);
   }
   \draw[blue,thick] (11.6,-.4) rectangle (12.4,5.4);
   \draw[blue,thick] (-.4,-.4) rectangle (.4,5.4);
   
   \begin{scope}[yshift=-9cm]
      \draw[-latex] (-1.5,-1) -- ++(3,0) node[anchor=north] {1};
      \draw[-latex] (-1,-1.5) -- ++(0,3) node[anchor=east,text width=12mm,align=right] {P dir.\\$k \neq 1$};
      
      \foreach \x in {0,1,...,5} {
         \foreach \y in {0,1,...,5} {
            \draw[fill=black] (\x,\y) circle (.1);
            \draw[thick] (\x,\y) -- ++(1,0);
            \draw[thick] (\x,\y) -- ++(0,1);
         }
      }
      \foreach \x in {6,7,...,11} {
         \foreach \y in {0,1,...,5} {
            \draw[black!40,fill=black!40] (\x,\y) circle (.1);
            \draw[thick,black!40] (\x,\y) -- ++(1,0);
            \draw[thick,black!40] (\x,\y) -- ++(0,1);
         }
      }
      
      \foreach \x in {0,1,...,5} {
         \draw[red,thick] (\x,6) circle (.1);
      }
      \draw[red,thick] (-.3,5.7) rectangle (5.3,6.3);
      \draw[red,thick] (-.3,-.3) rectangle (5.3,.3);
      
      \foreach \x in {6,7,...,11} {
         \draw[green!70!black,thick] (\x,6) circle (.1);
      }
      \draw[green!70!black,thick] (5.7,5.7) rectangle (11.3,6.3);
      \draw[green!70!black,thick] (5.7,-.3) rectangle (11.3,.3);
      
      \foreach \y in {0,1,...,5} {
         \draw[blue,thick] (12,\y) circle (.1);
      }
      \draw[blue,thick] (11.6,-.4) rectangle (12.4,5.4);
      \draw[blue,thick] (-.4,-.4) rectangle (.4,5.4);
   \end{scope}
   
\end{tikzpicture}
\caption{%
        Global geometry of extended lattice. The \textit{top diagram}
        represents a section of the extended lattice along a $(1,k)$ plane
        where $k=2,3$ is a direction with \Cstar boundary conditions. All
        fields are periodic along the extended direction 1. \Cstar boundary
        conditions in the direction $k=2,3$ are replaced by shifted boundary
        conditions in the extended lattice. Shifted boundary conditions are
        imposed by properly defining the nearest neighbours of boundary sites.
        Empty circles in the red (resp. green, blue) rectangle have to be
        identified with the corresponding solid circles in the red (resp.
        green, blue) rectangle. The \textit{bottom diagram} represents a
        section of the extended lattice along a $(1,k)$ plane where $k=2,3$ is
        a periodic direction.  In \textit{both diagrams}, the black circles
        represent the sites of the physical lattice, and the grey circles
        represent the sites of the mirror lattice.
       }\label{fig:cstar}
\end{figure}
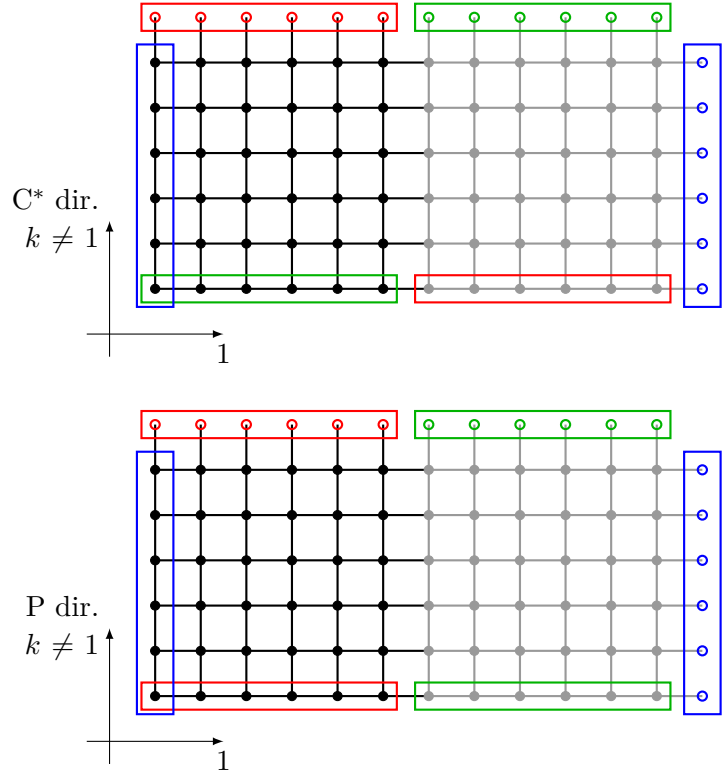

Other than the variety of boundary conditions in the temporal direction
inherited from \oQCD[-1.6], the \oQxD code allows for periodic or \Cstar
boundary conditions to be chosen in the spatial directions.  If the gauge
fields satisfy periodic boundary conditions in \textit{all} spatial directions
$k$, the fermion fields $\psi_f(x)$ and $\bar{\psi}_f(x)$ satisfy general
phase--periodic boundary conditions ($f$ is the flavour index), i.e.
\begin{align}\label{eq:periodic-bc-gauge}
   U_\mu(x + L_k \hat{e}_k) &= U_\mu(x)                                  \;, &
   A_\mu(x + L_k \hat{e}_k) &= A_\mu(x)                                  \;, \\\label{eq:periodic-bc-quark}
   \psi_f(x + L_k \hat{e}_k) &= e^{i\theta_{f,k}} \psi_f(x)              \;, & 
   \bar{\psi}_f(x + L_k \hat{e}_k) &= e^{-i\theta_{f,k}} \bar{\psi}_f(x) \;.
\end{align}
Phase--periodic boundary conditions are incompatible with \Cstar{} boundary
conditions. If the gauge fields satisfy \Cstar{} boundary conditions in at least
one direction, say $k$, then $\theta_{f,j}=0$ for all $f$ and $j$, and
\begin{align}\label{eq:cstar-bc-gauge}
   U_\mu(x + L_k \hat{e}_k)  &= U_\mu^*(x)               \;, &
   A_\mu(x + L_k \hat{e}_k)  &= -A_\mu(x)                \;, \\\label{eq:cstar-bc-quark}
   \psi_f(x + L_k \hat{e}_k) &= C^{-1} \bar{\psi}_f^T(x) \;, &
   \bar{\psi}_f(x + L_k \hat{e}_k) &= -\psi_f^T(x) C     \;.
\end{align}
The charge--conjugation matrix $C$ satisfies
\begin{align}
  C^T                 &= -C \;, & 
  C^\dag              &=  C^{-1} \;,  &
  C^{-1} \gamma_\mu C &= - \gamma_\mu^T \;.
\end{align}

\Cstar boundary conditions are implemented by means of an orbifold
construction. Assume that $k=1$ is a direction with \Cstar boundary
conditions,%
\footnote{%
         In the input file of a typical main program in \oQxD (see
         section~\ref{sec:iso1}), one can choose the number of spatial
         directions with \Cstar boundary conditions. \Cstar boundary conditions
         are turned on sequentially in directions 1, 2 and 3.
         }
in order to simulate a \textit{physical lattice} with size $V = L_0 \times
{L}_1 \times {L}_2 \times {L}_3$ the \texttt{openQ*D} code allocates a lattice
with size $\VCs = L_0 \times (2 {L}_1) \times {L}_2 \times {L}_3$, which
we will refer to as the \textit{extended lattice}.  Points in the
\textit{physical lattice} are assumed to have coordinates which satisfy $0 \le
x_\mu < {L}_\mu$. The extended lattice can be interpreted as a double--covering
of the physical lattice, with coordinates satisfying $0 \le x_\mu < {L}_\mu$ for
$\mu \neq 1$ and $0 \le x_1< 2{L}_1$. Points outside the physical lattice
constitute the \textit{mirror lattice}. On the extended lattice, points $x$ and
$x + L_k \hat{e}_k$ do not coincide, so eqs. \eqref{eq:cstar-bc-gauge} and
\eqref{eq:cstar-bc-quark} have to be interpreted as constraints which define the
admissible gauge and fermion fields. These are referred to as the
\textit{orbifold constraints}. While the admissible gauge fields in the mirror
lattice are completely determined by the value of the gauge field in the
physical lattice via \eqref{eq:cstar-bc-gauge}, the orbifold constraint has a
different meaning for fermion fields, providing a relation between $\psi$ in the
physical lattice and $\bar{\psi}$ in the mirror lattice, and vice versa. Given
that the fermion fields $\psi$ and $\bar{\psi}$ are independent Grassmanian
variables on the physical lattice, then one can equivalently choose the value of
$\psi$ in each point of the extended lattice as a complete set of independent
variables. The integration of the Grassmanian variables yields the Pfaffian of
the operator $CTD$~\cite{Lucini:2015hfa}, where $T$ is the translation operator
defined by
\begin{gather}
   T\psi(x) = \psi(x+L_1 \hat{e}_1) \;.
\end{gather}
One easily proves that
\begin{gather}
   \left| \text{pf} \, ( CTD ) \right| = \left| \det D \right|^{1/2} \;,
\end{gather}
which justifies the need for $\alpha_f=1/4$ in eq.~\eqref{eq:physdistr}.
Since the square of the charge–conjugation operation is the identity, all
fields must obey  periodic boundary conditions along the extended direction
$k=1$, i.e.
\begin{align}\label{eq:ext-periodic-bc-gauge}
   U_\mu(x + 2L_1 \hat{e}_1) &= U_\mu(x)                \;, &
   A_\mu(x + 2L_1 \hat{e}_1) &= A_\mu(x)                \;, \\\label{eq:ext-periodic-bc-quark}
   \psi_f(x + 2L_1 \hat{e}_1) &= \psi_f(x)              \;, &
   \bar{\psi}_f(x + 2L_1 \hat{e}_1) &= \bar{\psi}_f(x)  \;.
\end{align}
\Cstar boundary conditions in directions $k = 2, 3$ are implemented by
modifying the global topology of the extended lattice (see
fig.~\ref{fig:cstar}). In fact in these directions, \Cstar boundary conditions
in the physical lattice imply shifted boundary conditions in the extended
lattice, i.e.
\begin{align}\label{eq:ext-shifted-bc-gauge}
   U_\mu(x + L_k \hat{e}_k) &= U_\mu(x + L_1 \hat{e}_1)                \;, &
   A_\mu(x + L_k \hat{e}_k) &= A_\mu(x + L_1 \hat{e}_1)                \;, \\\label{eq:ext-shifted-bc-quark}
   \psi_f(x + L_k \hat{e}_k) &= \psi_f(x + L_1 \hat{e}_1)              \;, &
   \bar{\psi}_f(x + L_k \hat{e}_k) &= \bar{\psi}_f(x + L_1 \hat{e}_1)  \;. 
\end{align}
When the determinant of the Dirac operator is stochastically estimated by means
of a pseudofermion action as in eq.~\eqref{eq:action}, the pseudofermion field
$\Phi_{f,\ell}$ is natively defined on the extended lattice, i.e.
$\Phi_{f,\ell}(x)$ are truly independent variables for each $x$ in the extended
lattice. Moreover it satisfies the same boundary conditions as $\psi_f$ in
eqs.~\eqref{eq:ext-periodic-bc-quark} and~\eqref{eq:ext-shifted-bc-quark}.

It is worth noticing that \Cstar boundary conditions can be implemented in
different ways. For instance, the implementation proposed
in appendix D of ref.~\cite{Lucini:2015hfa} does not double the lattice, but the number of
pseudofermion fields. Roughly speaking one needs to represent quarks and
antiquarks by means of independent pseudofermion fields which are mixed by the
boundary conditions. The \oQxD implementation simply maps each pair of
pseudofermion fields in the geometry of the extended lattice. The cost of the
application of the Dirac operator implemented as in \oQxD and as
in~\cite{Lucini:2015hfa} is exactly identical.  Therefore, as far as the
application and inversion of the Dirac operator, the orbifold contruction does
not introduce any overhead with respect to more standard implementations of
\Cstar boundary conditions. On the other hand, the gauge field is evolved
twice. In principle one could evolve the gauge field only on the physical
lattice and then copy its value to the mirror lattice. This strategy will be
considered in the future. However, simulations close to the physical point are
dominated by the inversion of the Dirac operator and the overhead due to the
evolution of the gauge field is expected to be negligible.  Evidence of this
fact has been presented in~\cite{Campos:2017fly}.

%% file: 24-definitions.tex
The SU(3) and compact U(1) gauge actions that can be simulated with \oQxD are 
\begin{gather}\label{eq:action:SU3}
   \SgSU = \frac{\omega_{\Cs}}{g_0^2} \sum_{k=0}^1 c^\mathrm{SU(3)}_k \sum_{\mathcal{C} \in \mathcal{S}_k} \tr [ 1 - U(\mathcal{C}) ] \;, \\\label{eq:action:U1}
   \SgU  = \frac{\omega_{\Cs}}{2 \qel^2 e_0^2} \sum_{k=0}^1 c^\mathrm{U(1)}_k \sum_{\mathcal{C} \in \mathcal{S}_k} [ 1 - z(\mathcal{C}) ]  \;, 
\end{gather}
where  $U(\mathcal{C})$ and $z(\mathcal{C})$ denote the SU(3) and U(1) parallel
transports along  a path $\mathcal{C}$ on the lattice.  $\mathcal{S}_0$ and
$\mathcal{S}_1$ are the sets of all oriented plaquettes and all oriented $1
\times 2$ planar loops respectively and the overall weight $\omega_{\Cs}$ is
$1$ if no \Cstar boundary conditions are used.  With \Cstar boundary conditions
$\omega_{\Cs}=1/2$ corrects for the double counting introduced by summing over
all plaquette and double--plaquette loops in the extended lattice instead of
the physical lattice (c.f. section~\ref{sec:cstar}). The coefficients $c_{0,1}$
satisfy the relation $c_0 + 8c_1 = 1$. For SU(3), the Wilson action is obtained by
choosing $c_0 = 1$, the tree--level improved Symanzik (or L\"uscher--Weisz)
action is obtained by choosing $c_0 = \tfrac{5}{3}$, and the Iwasaki action is
obtained by choosing $c_0 = 3.648$.  The parameters $g_0$ and $e_0$ are the bare SU(3)
and U(1) gauge couplings respectively, which are related to the $\beta$ parameter and the bare fine--structure constant $\alpha_0$ by
\begin{gather}
   \beta = \frac{6}{g_0^2} \;, \qquad
   \alpha_0 = \frac{e_0^2}{4\pi} \;.
\end{gather}
In the compact formulation of QED, all electric charges must be integer
multiples of some elementary charge $\qel$ which is defined in units of the
charge of the positron. As discussed in ref.~\cite{Lucini:2015hfa}, $\qel$
appears as an overall factor in the gauge action and essentially sets the
normalization of the U(1) gauge field in the continuum limit. Even though in
infinite volume $\qel=1/3$ would be an appropriate choice in order to simulate
quarks, in finite volume with \Cstar boundary conditions one needs to choose
$\qel=1/6$ in order to construct gauge--invariant interpolating operators for
charged hadrons~\cite{Lucini:2015hfa,Hansen:2018zre}.  Note that by using a
compact formulation of QED, no gauge fixing is added to the action, and
furthermore the user is free to choose simulating (QCD+)QED without \Cstar
boundary conditions.

The actions in eqs.~\eqref{eq:action:SU3} and~\eqref{eq:action:U1} assume
periodic boundary conditions in time. In the more general case, the actions are
modified at the time boundary in order to allow for $O(a)$ improvement. The
general form of the gauge actions can be found in~\cite{oQxDdocu:gauge_action}.

%% file: 25-dirac.tex
The Dirac operator implemented in \oQxD is given by a sum of terms
\begin{gather}\label{eq:dirac}
        D = m_0 + \D[w] + \delta \D[sw] + \delta \D[b]  \;,
\end{gather}
where $\D[w]$ is the (unimproved) Wilson--Dirac operator, $\delta\D[sw]$ is the
Sheikholeslami--Wohlert (SW) term, and $\delta\D[b]$ is the time boundary
$O(a)$--improvement term. For simplicity, periodic boundary conditions in the
time direction will be assumed, which means $\delta\D[b]=0$. The definition of
$\delta\D[b]$ for other boundary conditions can be found in~\cite{oQxDdocu:dirac}.
The Wilson--Dirac operator of eq.~\eqref{eq:dirac} can be written as
\begin{gather}\label{eq:wilson-dirac}
   \D[w] = \sum_{\mu=0}^3 \frac{1}{2} \left\{ \gamma_\mu ( \nabla_\mu + \nabla^*_\mu ) - \nabla^*_\mu \nabla_\mu \right\} \;,
\end{gather}
where the covariant derivatives are defined as
\begin{align}\label{eq:frw-covder}
   \nabla_\mu   \psi(x) &= U(x,\mu) z(x,\mu)^{\hat{q}} \psi(x+\hat{\mu}) - \psi(x) \;, \\\label{eq:bkw-covder}
   \nabla^*_\mu \psi(x) &= \psi(x) - U(x-\hat{\mu},\mu)^\dag z(x-\hat{\mu},\mu)^{-\hat{q}} \psi(x-\hat{\mu})  \;.
\end{align}
The SW term is given by
\begin{gather}\label{eq:sw-term}
   \delta\D[sw] =
          \cswSU \sum_{\mu,\nu=0}^3 \tfrac{i}{4} \sigma_{\mu\nu} \Fhat_{\mu\nu}
   + q \, \cswU  \sum_{\mu,\nu=0}^3 \tfrac{i}{4} \sigma_{\mu\nu} \Ahat_{\mu\nu}  \;.
\end{gather}
The SU(3) field tensor $\Fhat_{\mu\nu}(x)$ and the U(1) field tensor
$\widehat{A}_{\mu\nu}(x)$ are constructed in terms of the clover plaquette. The
explicit expression of the SU(3) field tensor used in \texttt{openQ*D} can be
found in ref.~\cite{Luscher:1996sc}, while the U(1) field tensor is given here,
\begin{align}
   \Ahat_{\mu\nu}(x) &= \tfrac{i}{4 q_\text{el}} \Im\left\{
   z_{\mu\nu}(x)+ 
   z_{\mu\nu}(x-\hat{\mu})+
   z_{\mu\nu}(x-\hat{\nu})+
   z_{\mu\nu}(x-\hat{\mu}-\hat{\nu})
   \right\}
   \ , \\
   z_{\mu\nu}(x) &= z(x,\mu) z(x+\hat{\mu},\nu) z(x+\hat{\nu},\mu)^\dag z(x,\nu)^\dag
   \ .
\end{align}
The normalization is chosen in such a way that $ -i e_0 \Ahat_{\mu\nu}(x) $ is
the canonically--normalized field tensor in the naive continuum limit. Notice
that the field tensors are anti--hermitian.

In presence of electromagnetism, the Dirac operator depends on the electric
charge of the quark field. Let $q$ be the physical electric charge in units of
$e$ (i.e. $q=2/3$ for the up quark, and $q=-1/3$ for the down quark). In the
compact formulation of QED, all electric charges must be integer multiples of
an elementary charge $\qel$, which appears as a parameter in the U(1) gauge
action~\eqref{eq:action:U1}. The integer parameter
\begin{gather}
   \hat{q} = \frac{q}{\qel} \in \mathbb{Z}
\end{gather}
is the one appearing in the hopping term in eqs.~\eqref{eq:frw-covder} and
\eqref{eq:bkw-covder}. On the other hand, notice that the SW
term~\eqref{eq:sw-term} is written in terms of the physical charge $q$. This
normalization corresponds to a definition of $\cswU$ which is equal to 1 at
tree level.
The definition of the even--odd preconditioned Dirac operator $\Dhat$ is
standard~\cite{DeGrand:1988vx}
\begin{gather}
   \hat{D} = D_\text{ee} - D_\text{eo} D_\text{oo}^{-1} D_\text{oe}
   \ , \hspace{1cm}
   D =
   \begin{pmatrix}
      D_\text{ee} & D_\text{eo} \\
      D_\text{oe} & D_\text{oo}
   \end{pmatrix}
   \ ,
\end{gather}
and so is the definition of the \textit{small-determinant} action $\Ssdet$
appearing in eq.~\eqref{eq:physdistr}
\begin{gather}
  \Ssdet = -\sum_{f} \alpha_f \,\tr \log  ({\mathbf{1}+ D_{f,\text{oo}}})
\ . 
\end{gather}

%% file: 30-userguide.tex
\subsection{Structure of the \oQxD program package}
\label{sec:structure}

The \oQxD code includes several main programs, roughly divided in three
categories: programs to generate configurations, programs to measure
observables, and utility programs. The following programs (in the \texttt{main}
directory) can be used to generate gauge configurations for various theories:
\begin{itemize}\setlength{\itemsep}{0em}
   \item \texttt{iso1}: SU(3)$\times$U(1) gauge theory with dynamical fermions;
   \item \texttt{qcd1}: SU(3) gauge theory with dynamical fermions;
   \item \texttt{ym1}: SU(3) pure gauge theory;
   \item \texttt{mxw1}: U(1) pure gauge theory.
\end{itemize}
The following programs (in the \texttt{main} directory) can be used to
calculate simple observables:
\begin{itemize}\setlength{\itemsep}{0em}
   \item \texttt{ms1}: reweighting factors (see section~\ref{sec:iso1} and
   appendix~\ref{sec:rhmc});
   \item \texttt{ms2}: spectral range of $(\Dhat^\dag\Dhat)^{1/2}$
   ($\Dhat$ is the even--odd preconditioned Dirac operator);
   \item \texttt{ms3}: SU(3) Wilson--flow observables;
   \item \texttt{ms4}: quark propagators;
   \item \texttt{ms5}: U(1) Wilson--flow observables;
   \item \texttt{ms6}: neutral pseudoscalar--pseudoscalar and axial--pseudoscalar correlators.
\end{itemize}
Finally, the following utility programs are also included:
\begin{itemize}\setlength{\itemsep}{0em}
   \item \texttt{minmax/minmax}: it generates the rational approximations needed
   for the RHMC algorithm;
   \item \texttt{devel/nompi/read*}: they can be used to read the binary
   \texttt{*.dat} files generated by the other programs.
\end{itemize}

\subsection{User guide for the dynamical QCD+QED simulation program \texttt{iso1}}
\label{sec:iso1}

\subsubsection{Compiling and running the main program}\label{sec:compilerun}

A complete guide to the usage of all programs listed in
section~\ref{sec:structure} can be found in the headers of the source--code
files, and in the \texttt{README} files in the corresponding directories. Often
the user will be referred to other sources of documentation (e.g.
\texttt{README} files in some of the \texttt{modules} subdirectories,
or the headers of other source--code files, and some of the PDF files in the
\texttt{doc} directory). This section is intended to be neither a replacement
nor a duplicate of these sources of documentation, but rather an overview of
the main steps that are needed to use the \texttt{iso1} program to generate
QCD+QED configurations.

\begin{enumerate}
   \item \textbf{Download the code and check the dependences.} The code is
   publicly available on GitLab at \url{https://gitlab.com/rcstar/openQxD}. The
   simulation and measurement programs, i.e. all programs in the \texttt{main}
   directory, require some MPI libraries compliant with the MPI 1.2 (or later)
   standard. The \texttt{minmax} program requires the GMP
   (\url{https://gmplib.org}) and GNU MPFR (\url{http://www.mpfr.org})
   libraries. Notice that the \texttt{minmax} program can be run on a personal
   computer and does not need MPI, therefore one does not need to install the
   GMP and GNU MPFR libraries on production machines.
   \item \textbf{Set the environment variables.} The Makefile in the
   \texttt{main} directory assumes that the C compiler can be called by using
   \texttt{\$(GCC)}, the MPI header file is found at
   \texttt{\$(MPI\_INCLUDE)/mpi.h}, the MPI compiled library is found in the
   \texttt{\$(MPI\_HOME)/lib/} directory, and the \texttt{mpicc} command is
   available. The needed environment variables can be defined in the appropriate
   shell initialization files, e.g.
\begin{lstlisting}[language=other]
#!/bin/bash
# [Stuff]

export GCC="gcc"
export MPI_INCLUDE="/usr/local/include/"
export MPI_HOME="/usr/local/"
\end{lstlisting}
   \item \textbf{Choose the intrinsics acceleration options.} Some pieces of
   code exist in several versions: plain C, inline--assembly with SSE
   instructions, and inline--assembly with AVX instructions. The default Makefile
   uses the C version of the code. In order to use the inline--assembly version,
   one needs to modify the \texttt{CFLAGS} variable defined in lines 122--124 of
   \texttt{main/Makefile}. For instance, on some x86-64 machines one can use
   
\begin{lstlisting}[language=other,numbers=left,firstnumber=122]
CFLAGS = -std=c89 -pedantic -fstrict-aliasing \
         -Wall -Wno-long-long -Wstrict-prototypes \
         -Werror -O -mno-avx -DAVX -DFMA3 -DPM
\end{lstlisting}
   which activates AVX and FMA3 instructions and assumes that prefetch
   instructions fetch 64 bytes at a time. For a full description of available
   options, refer to the \texttt{README} file in the root directory.
   \item \textbf{Choose the lattice geometry.} The lattice geometry is chosen at
   compile time by modifying the macros defined in the first part of the
   \texttt{include/global.h} file. A full description of these macros can be
   found in the \texttt{main/README.global} file. For instance the following
   choice
\begin{lstlisting}[language=other,numbers=left,firstnumber=18]
#define NPROC0 8
#define NPROC1 8
#define NPROC2 4
#define NPROC3 4

#define L0 8
#define L1 8
#define L2 8
#define L3 8
\end{lstlisting}
   corresponds to an $8^4$ local lattice, replicated on an $8^2 \times 4^2$ MPI
   process grid (the code will need to be run with 1024 MPI processes), which
   yields a $64^2 \times 32^2$ global lattice. As explained in
   section~\ref{sec:cstar}, this choice of simulation parameters 
   corresponds to a $64^2 \times 32^2$ physical global lattice if no \Cstar
   boundary conditions are used, or to a $64 \times 32^3$ physical global
   lattice if \Cstar boundary conditions are used in at least one spatial
   direction. In our implementation, \texttt{NPROCn} has to be a
   multiple of 2 if \Cstar boundary conditions are used in the direction
   $\texttt{n}=1,2,3$.
   \item \textbf{Compile the \texttt{iso1} program and prepare for running.} At
   this point, the code is ready to be compiled. Assuming that the root
   directory of the code is \texttt{\${HOME}/openQxD}, this is done by executing
   the following commands in a bash shell.
\begin{lstlisting}[language=other]
cd ${HOME}/openQxD/main
make iso1
\end{lstlisting}
   One can set up the directories and files to run the code by executing the
   following commands in a bash shell.
\begin{lstlisting}[language=other]
cd ${HOME}/openQxD
mkdir test
cd test
mkdir cnfg dat log input
cp ../main/iso1 iso1
> input/pedro01.in
> runtest.sh
chmod a+x runtest.sh
\end{lstlisting}
   \item \textbf{Edit the input file.} The input file \texttt{input/pedro01.in}
   must contain all adjustable parameters of the simulation (except the few ones
   that have been set at compile time). A rough guide on how to construct an
   input file for the \texttt{iso1} program is found in
   section~\ref{sec:input}. Alternatively, a sample input file can be cut
   and paste from appendix~\ref{app:pedro01.in}.
   
   \item \textbf{Start the simulation.} Edit the \texttt{runtest.sh} script as
   follows:
\begin{lstlisting}[language=other]
#!/bin/bash
./iso1 -i input/pedro01.in -noloc -rmold
\end{lstlisting}
   The \texttt{runtest.sh} script contains the command that invokes the
   \texttt{iso1} program. It can be launched via a standard \texttt{mpirun}
   command, or incorporated in a script for a job scheduler. Recall that the
   number of needed MPI processes has been decided at compile time, and it is
   equal to 1024 in this case. The \texttt{iso1} program takes a number of
   command--line options: the input file is specified with the \texttt{-i}
   option, the \texttt{-noloc} option specifies that the configuration files
   must be saved by a single MPI process, the \texttt{-rmold} specifies that
   only the most recent configuration must be kept and all previous ones must
   be deleted. The program will start the simulation from a randomly generated
   configuration. More details about the command--line options can be found in
   the \texttt{main/README.iso1} file.

   \item \textbf{Interrupt the simulation.} Assuming that no error is produced,
   the simulation code will end naturally when all the configurations requested
   in the input file are generated. If the simulation needs to be interrupted
   earlier, one can just execute the following commands in a bash shell.
\begin{lstlisting}[language=other]
cd ${HOME}/openQxD/test
touch log/pedro01.end
\end{lstlisting}
   The simulation code will stop gracefully right after the next configuration
   is saved.
   
   \item \textbf{Resume the simulation.} Assuming that the last generated
   configuration was \texttt{pedro01n42}, edit the input file and set the
   \texttt{nth} variable in the \texttt{[MD trajectories]} section to 0 (see
   below for a description of the input file), and edit the \texttt{runtest.sh}
   script as follows:
\begin{lstlisting}[language=other]
#!/bin/bash
./iso1 -i input/pedro01.in -noloc -rmold -c pedro01n42 -a
\end{lstlisting}
   Once this is executed, the simulation will continue from where it was
   interrupted.

\end{enumerate}

\subsubsection{Constructing the input file for \texttt{iso1}}\label{sec:input}

Most of the parameters needed to generate configurations are passed to the
\texttt{iso1} program by means of a human--readable input file, in this case
\texttt{pedro01.in} in the \texttt{test/input} directory. For a full
description of the various parameters, the reader is referred to the
\texttt{main/README.iso1} and \texttt{doc/parms.pdf} files (and references
therein). A rough guide to the various sections that compose the input file is
provided here, with no ambition of completeness.
\begin{enumerate}

   \item \textbf{Run name and output directories.}
\begin{lstlisting}
[Run name]
name       pedro01

[Directories]
log_dir    ./log      # absolute path, or relative path
dat_dir    ./dat      # to the working directory of iso1
cnfg_dir   ./cnfg
\end{lstlisting}
   The program \texttt{iso1} will produce several output files:
   \begin{itemize}\setlength{\itemsep}{0em}
      \item \texttt{./log/pedro01.log}, human--readable file, with general
      information about the simulation;
      \item \texttt{./dat/pedro01.dat}, binary file, with the history of simple
      diagnostic observables;
      \item \texttt{./dat/pedro01.ms3.dat} and \texttt{./dat/pedro01.ms5.dat},
      binary files, with the history of SU(3) and U(1) Wilson flow observables;
      \item \texttt{./dat/pedro01.par}, binary file, with all simulation
      parameters;
      \item \texttt{./dat/pedro01.rng}, binary file, with the state of the
      random number generator at the time of the most recent saved
      configuration;
      \item \texttt{./cnfg/pedro01n*}, binary files, with the gauge
      configuration.
   \end{itemize}
   For every file in the \texttt{log} and \texttt{dat} directories, a backup
   file identified by a tilde at the end of its name is created and updated
   every time a configuration is saved.
   
   \item \textbf{Schedule management.}
\begin{lstlisting}
[MD trajectories]
nth         100       # multiple of dtr_cnfg
ntr         800       # multiple of dtr_cnfg
dtr_log     5
dtr_ms      10        # multiple of dtr_log
dtr_cnfg    50        # multiple of dtr_ms
\end{lstlisting}
   The program \texttt{iso1} will print one entry in the log file every 5 MD
   trajectories, will measure and print Wilson flow observables every 10 MD
   trajectories, will save a configuration every 50 MD trajectories. The first
   100 trajectories are considered of thermalization (no observables are
   measured), a total of 800 MD trajectories will be generated and 15
   configurations will be saved.

   \item \textbf{Ranlux \cite{Luscher:1993dy} initialization.}
\begin{lstlisting}
[Random number generator]
level      0          # this should not be changed
seed       19521      # this can be any positive integer
\end{lstlisting}

   \item \textbf{Boundary conditions.}
\begin{lstlisting}
[Boundary conditions]
type       periodic   # or SF, open, open-SF
cstar      3          # or 0, 1, 2
\end{lstlisting}
   In this case periodic boundary conditions are chosen in time, and \Cstar
   boundary conditions in all 3 spatial directions. The implementation of
   \Cstar boundary conditions in \oQxD is described in section~\ref{sec:cstar}.
   If SF or open--SF boundary conditions are chosen in time, the number of
   parameters in this section increases, as one needs to specify the value of
   the fields on the SF boundaries. For a full description of these parameters,
   refer to \texttt{doc/parms.pdf}.
   
   \item \textbf{Gauge actions.}
\begin{lstlisting}
[SU(3) action]
beta       5.3
c0         1.0      # 1=Wilson, 5/3=Lüscher-Weisz, 3.648=Iwasaki

[U(1) action]
type       compact  # only option currently available
alpha      0.05     # bare fine-structure constant
invqel     6.0      # see "Quark flavours" below 
c0         1.0      # Wilson action
\end{lstlisting}
   If different boundary conditions in time are chosen, the number of parameters
   in these sections increases, as one needs to specify the $O(a)$--improvement
   boundary coefficients. Refer to \texttt{doc/gauge\_action.pdf},
   \texttt{doc/parms.pdf} of all these parameters.
   
   \item \textbf{Quark flavours.} In the terminology of the \oQxD
   code, a \textit{quark flavour} is identified by all adjustable parameters
   that define the Dirac operator. For instance, in a simulation in the isospin
   symmetric limit, the up and down quark count as a single quark flavour. In
   the following example, two quark flavours are requested, and the parameters
   of the corresponding Dirac operators are initialized.
\begin{lstlisting}
[Quark action]
nfl        2

[Flavour 0]          # Down quark
qhat       -2        # qhat must be integer
                     # el. charge = qhat/invqel = -2/6 = -1/3
kappa      0.136377  # hopping parameter
su3csw     1.909520  # u1csw=su3csw=0 => no O(a) improv.
u1csw      1.0       # u1csw=su3csw=1 => tree-level O(a) improv.

[Flavour 1]          # Up quark
qhat       4         # el. charge = qhat/invqel = 4/6 = 2/3
kappa      0.137312
su3csw     1.909520
u1csw      1.0
\end{lstlisting}
   If different boundary conditions in time are chosen, the number of parameters
   in these sections increases, as one needs to specify the $O(a)$--improvement
   boundary coefficients. Also, if no \Cstar boundary conditions are used,
   one can choose phase--periodic boundary conditions for fermions in space.
   Refer to \texttt{doc/dirac.pdf}, \texttt{doc/parms.pdf} for a detailed
   explanation of all these parameters.
   
   \item \textbf{Rational approximation.} With \Cstar boundary conditions,
   the Pfaffian of the even--odd preconditioned Dirac operator $\hat{D}$ is
   needed, whose absolute value can be generated by a pseudofermion effective
   action of the type $\psi^\dag (\Dhat^\dag \Dhat)^{-1/4} \psi$. The
   fractional power of $\Dhat^\dag \Dhat$ is replaced by a rational
   approximation, which must be generated by means of the \texttt{minmax}
   program~\cite{zbMATH03014451,Ralston1965}. We sketch here how to use this program, see
   \texttt{minmax/README} for more details.
   
   First, one needs to modify the \texttt{GCC} and \texttt{MPLIBPATH} variables
   in \texttt{minmax/Makefile}. The Makefile assumes that the C compiler can be
   called by using \texttt{\$(GCC)}, the GMP and MPFR header files are found in
   the \texttt{\$(MPLIBPATH)/include/} directory, and the compiled libraries are
   found in the \texttt{\$(MPLIBPATH)/lib/} directory.
\begin{lstlisting}[language=other,numbers=left,firstnumber=22]
GCC = gcc

MPLIBPATH = /usr/local
\end{lstlisting}
   
   The \texttt{minmax} program is compiled and executed with the following
   commands in a bash shell.
\begin{lstlisting}[language=other]
cd ${HOME}/openQxD/minmax
make
./minmax -p -1 -q 4 -ra 1.98e-03 -rb 7.62 -goal 6e-05
\end{lstlisting}
   A rational approximation for $(\Dhat^\dag \Dhat)^\alpha$ is requested,
   with $\alpha=(-1)/(4)$ (\texttt{-p} and \texttt{-q} options), assuming that
   the eigenvalues of $(\Dhat^\dag\Dhat)^{1/2}$ are in the interval $[1.98
   \times 10^{-3} , 7.62]$ (\texttt{-ra} and \texttt{-rb} options), with a
   target relative precision of $6 \times 10^{-5}$ (\texttt{-goal} option). The
   spectral range of $(\Dhat^\dag\Dhat)^{1/2}$ must be guessed at first, but
   after some configurations have been generated it can be calculated with the
   program \texttt{main/ms2}. The \texttt{minmax} program creates a directory
   with a very long name, in this case\\[0.1em]
   \hspace*{1cm}\texttt{p-1q4mu0.00000000e+00ra1.98000000e-03rb7.62000000e+00}\\[0.1em]
   which contains several files named \texttt{n*.in}. The integer in the file
   name corresponds to the order of the generated rational approximation. Only
   the highest order rational approximation, \texttt{n10.in} in this case, meets
   the requested precision. The full content of the \texttt{n10.in} must be
   pasted in the input file in a section of the following type,
\begin{lstlisting}
[Rational 0]
power      -1 4
degree     10
range      1.98000000e-03 7.62000000e+00
mu         0.00000000e+00
delta      5.9691841082503071e-05
A          2.04978213590663732591e-01
nu[0]      1.22647978559899293316e+01
mu[0]      8.40737261524814627478e+00
# [...] the full content of n10.in must be pasted here
\end{lstlisting}
   Notice that more than one rational approximation can be used in the same
   input file (e.g. one may want to use different rational approximations for
   the up, down and strange quarks). Each rational approximation is identified
   by the integer in the section title.

   \item \textbf{MD Hamiltonian and integrator.}
\begin{lstlisting}
[HMC parameters]
actions    0 1 2 3  # List of action IDs, see below
npf        2        # Number of pseudofermions to be allocated
nlv        2        # Number of levels of integrator for MD eqs
tau        2.0      # MD trajectory length
facc       1        # Fourier acceleration for U(1) MD
                    # (0=not active, 1=active)

[Level 0]           # Innermost level
integrator OMF4     # Omelyan-Mryglod-Folk 4th order
nstep      2        # Number of times the elementary integrator
                    # is applied at this level
forces     0 1      # List of force IDs to be integrated at
                    # this level, see below

[Level 1]           # Outermost level
integrator OMF4
nstep      1
forces     2 3
\end{lstlisting}
   The MD Hamiltonian is given by the canonical kinetic term of the SU(3) gauge
   field, the kinetic term of the U(1) gauge field, and a sum of terms which do
   not depend on the MD momenta and are referred to as \textit{actions}. The
   kinetic term of the U(1) gauge field can be chosen to be of two types: the
   canonical one (\texttt{facc=0}), or the Fourier--accelerated one
   (\texttt{facc=1}). Refer to \texttt{doc/fourier.pdf} and
   section~\ref{sec:tech} for details on Fourier acceleration. The MD equations
   are solved by means of an approximate symplectic multilevel integrator,
   built in terms of standard elementary integrators. For each level, one needs
   to specify how many times the elementary integrator needs to be applied and
   which \textit{forces} need to be integrated. Refer to \texttt{doc/parms.pdf}
   and \texttt{module/update/README.mdint} for details on the integrator.

   The actions and forces are uniquely identified by an ID. Obviously there is a
   one--to--one correspondence between actions and forces. Corresponding actions
   and forces must share the same ID. The gauge actions and forces must be
   included, i.e.
\begin{lstlisting}
[Action 0]          # No adjustable parameters here!
action     ACG_SU3

[Force 0]
force      FRG_SU3

[Action 1]
action     ACG_U1

[Force 1]
force      FRG_U1
\end{lstlisting}
   
   In this example, two pseudofermion actions are used (notice that this number
   matches the number of pseudofermion fields requested in the \texttt{[HMC
   parameters]} section), one for \textit{up} quark and one for the
   \textit{down} quark.
\begin{lstlisting}
[Action 2]
action  ACF_RAT_SDET # Rational approximation effective action
ipf     0            # Pseudofermion ID (a number from 0 to 1)
ifl     0            # Flavour ID (down quark)
irat    0 0 9        # Use the rational approximation with ID=0
                     # Include all rat. appr. factors, 0 -> 9,
                     # i.e. no frequency splitting
isp     0            # Solver ID, used to generate the p.f. at
                     # the beginning of the MD and to calculate
                     # the Hamiltonian at the end of the MD

[Force 2]
force   FRF_RAT_SDET
isp     1            # Solver ID, used to calculate the force

[Action 3]
action  ACF_RAT_SDET
ipf     1            # Different pseudofermion ID
ifl     1            # Different flavour ID (up quark)
irat    0 0 9
isp     0

[Force 3]
force   FRF_RAT_SDET
isp     1
\end{lstlisting}
   Notice that \oQxD allows for frequency splitting (not used in this example):
   the poles and zeroes of the rational approximations can be separated in
   different pseudofermion actions. This is convenient because one may want to
   integrate different poles and zeroes in different levels of the integrator,
   and also one may want to use different solvers for different poles. For
   details on the pseudofermion actions and forces, and on the frequency
   splitting, one should refer to \texttt{doc/rhmc.pdf} and
   section~\ref{sec:tech}. 

   \item \textbf{Solvers.} Two multi--shift CG solvers are used in this example,
   with different residue for the actions and the forces.
\begin{lstlisting}
[Solver 0]
solver     MSCG        # or CGNE, SAP_GCR, DFL_SAP_GCR
nmx        2048        # Maximum number of iterations
res        1.0e-11     # Residue

[Solver 1]
solver     MSCG
nmx        2048
res        1.0e-8
\end{lstlisting}
   For details on the usage of other solvers, one should refer to
   \texttt{doc/parms.pdf}. The deflated solver (\texttt{DFL\_SAP\_GCR}) requires
   to set parameters for the generation and update of the deflation subspaces,
   also described in \texttt{doc/parms.pdf}. See also section~\ref{sec:dfltests}.
   
   \item \textbf{Wilson flow parameters.} The \texttt{iso1} program measures on
   the fly a number of simple observables (actions, SU(3) topological charge, electromagnetic
   fluxes) at positive flow time.
\begin{lstlisting}
[Wilson flow]
integrator RK3       # EULER: Euler, RK2: 2nd order Runge-Kutta
                     # RK3: 3rd order Runge-Kutta
eps        2.0e-2    # Integration step size
nstep      700       # Total number of integration steps
dnms       5         # Number of steps between measurements
\end{lstlisting}
   
\end{enumerate}

%% file: 40-performance.tex
For future reference and comparison, benchmark measurements have been performed
for the timing of the application of the double precision Wilson--Dirac operator
and the SAP (Schwartz--Alternating--Procedure) preconditioner. The HPC cluster at CERN has
been used, which features 72 nodes, each of them with two 8-core
Intel\textsuperscript{\textregistered} Xeon processors (E5-2630 v3, Haswell)
running at about 2.4\,GHz base frequency (3.6\,GHz max.). Nodes are connected
with Mellanox\textsuperscript{\textregistered} Infiniband FDR (56\,Gb/s). 

\begin{figure}[t]
 \centering
 {\includegraphics[width=0.49\textwidth,clip,page=1]{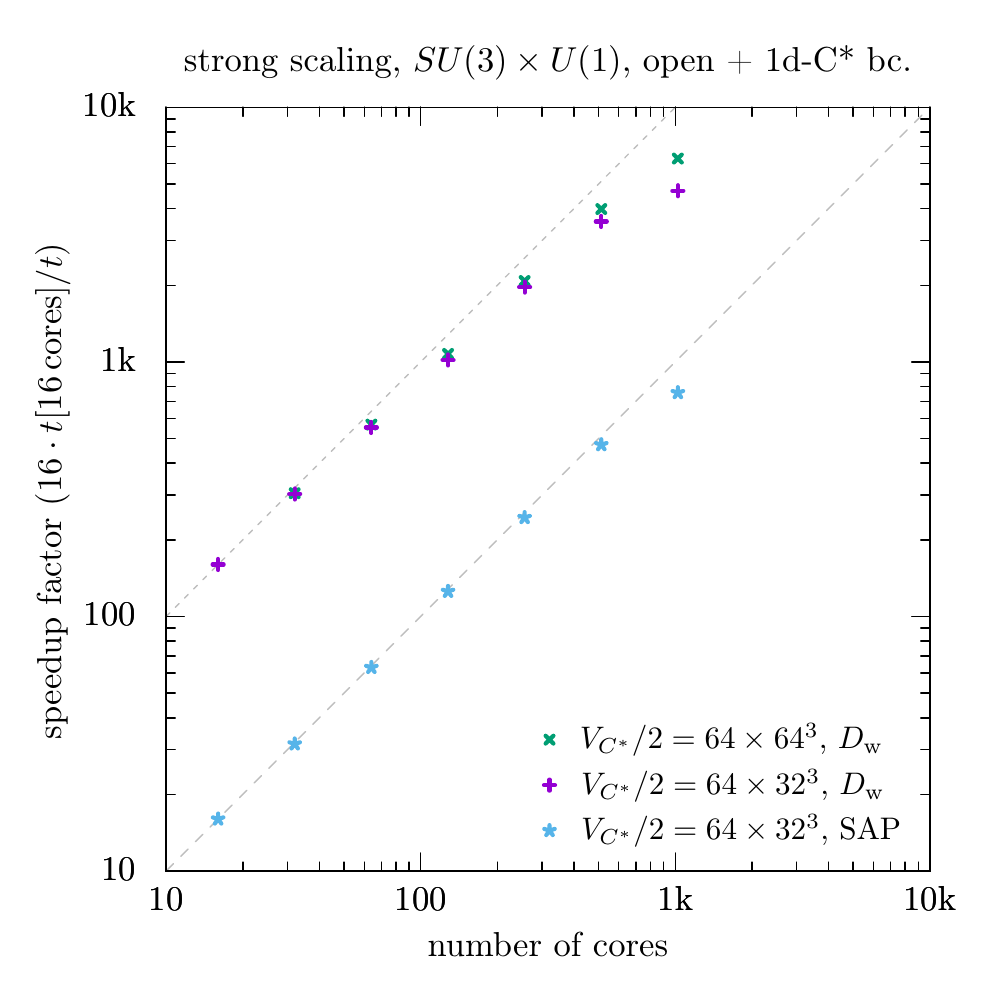}}
 {\includegraphics[width=0.49\textwidth,clip,page=2]{figures/scaling_all.pdf}}
  \caption{%
          Results for strong (left) and weak (right) scaling of the application
          of the Dirac operator and SAP preconditioner as explained in the
          text.  The speedup factors for the Dirac operator are multiplied by a
          factor 10 for better visibility. The dashed lines indicate perfect
          scaling behaviour accordingly.
          }\label{fig:scaling}
\end{figure}

The timings are obtained with the \texttt{time2} programs located in the
subdirectories \texttt{devel/dirac} and \texttt{devel/sap}.
All measured times have been normalised to the smallest partition (one node or
16 cores). The results of these scaling tests are shown in
fig.~\ref{fig:scaling}. A QCD+QED setup with open boundary conditions in time
and \Cstar boundary conditions in one spatial direction has been used.

The weak scaling test has been performed with a local lattice size of $8 \times
16 \times 8 \times 8$, giving an extended lattice with total volume $\VCs=2
\Nproc 8^4$. Because of the \Cstar boundary conditions this corresponds to a
physical lattice with volume $V=\Nproc 8^4$, cf. section~\ref{sec:cstar}. While
for the Dirac operator, parameters similar to the \textit{Quark flavours}
example (point 6) in section~\ref{sec:iso1} have been used, the SAP preconditioner
specifically employs a block size of $4^4$ with five SAP cycles (\texttt{ncy 5})
and five iterations (\texttt{nmr 5}) of the even--odd preconditioned Minimal
Residue (MinRes) block solver. The setup is similar for the strong scaling study
but with a constant total volume of $\VCs=2 \cdot 64\times 32^3$ and varying
local lattice sizes. In case of the double precision Wilson--Dirac operator, a
much larger lattice volume with $\VCs=2\cdot 64^4$ total lattice points was
probed as well. As it can be seen in the left panel of fig.~\ref{fig:scaling}
the larger lattice is performing even better than the smaller one.

In summary, the overall scaling studied here is close to optimal and small
deviations may partly result from remaining indigestions of the underlying
network.  Similar studies have to be done on other machines but the overall
behaviour is expected to be similar to the original \oQCD code.

%% file: 42-tests.tex
The \texttt{openQ*D} code has been tested by means of an extensive battery of
check programs, which can be found in the subdirectories of
\texttt{devel}.\footnote{The \texttt{devel} directory contains 46'224 lines of
code, against 60'203 lines of code in the \texttt{module} directory.} These
programs have been taken over from \texttt{openQCD-1.6} and \texttt{NSPT-1.4},
and extended in order to test the specific feature of the \texttt{openQ*D} code.
Roughly speaking, the check programs in each \texttt{devel} subdirectory test
features of the corresponding \texttt{module} subdirectory. Many check programs
test also interactions between different modules. These programs are meant to be
used by developers only and contain very limited documentation. Providing a
description of the check programs is outside of the scope of this paper, and a
short description can be found in the \texttt{INDEX} files in each
\texttt{devel} subdirectory. However, it is worth to point out a few facts. All
check programs have been run with all possible combinations of boundary
conditions in the space and temporal directions. Whenever possible, all
check programs have been run in a pure QCD setup (i.e. only the SU(3) gauge
field is allocated), a pure QED setup (i.e. only the U(1) gauge field is
allocated), and a QCD+QED setup (i.e. both gauge fields are allocated). All
check programs have been run with various geometric configurations, i.e. lattice
size and processor grid. Besides a plethora of minor details, specific check
programs have been written to test:
\begin{itemize}
   \item the implementation of \Cstar{} boundary conditions for both gauge
   fields and for the Dirac operator;
   \item general properties of the Dirac operator with generic electric charge
   (e.g. gauge convariance, translational covariance, $\gamma_5$--hermiticity,
   comparison to analytic expression in case of zero gauge field);
   \item the rational approximation of generic powers, and the associated
   reweighting factors;
   \item the forces for the U(1) field, the QED action, the U(1) Wilson flow,
   the U(1) observables (e.g. clover field tensor, electromagnetic fluxes);
   \item the MD with the U(1) field, with and without Fourier acceleration.
\end{itemize}

%% file: 44-hamiltonian.tex
The use of Fourier Acceleration in QCD+QED simulations modifies the MD
Hamiltonian and, consequently, the MD equations. In order to test the
consistency between the two, one can look at the violation $\Delta H$ of
Hamiltonian conservation as a function of the MD integration step--size $\Delta
\tau$. The violation should vanish as a positive power of the integration
step--size in the $\Delta \tau \to 0$ limit. The power depends on the chosen
integrator. When the total trajectory length is kept constant, the leap--frog
integrator (LF) and 2nd order Omelyan--Mryglod--Folk (OMF2) integrators yield
$\Delta H \sim (\Delta \tau)^2$, while the 4th order Omelyan--Mryglod--Folk (OMF4)
integrator yields $\Delta H \sim (\Delta \tau)^4$.

\begin{figure}[t]
   \small\centering
   \includegraphics[width=0.9\textwidth,clip]{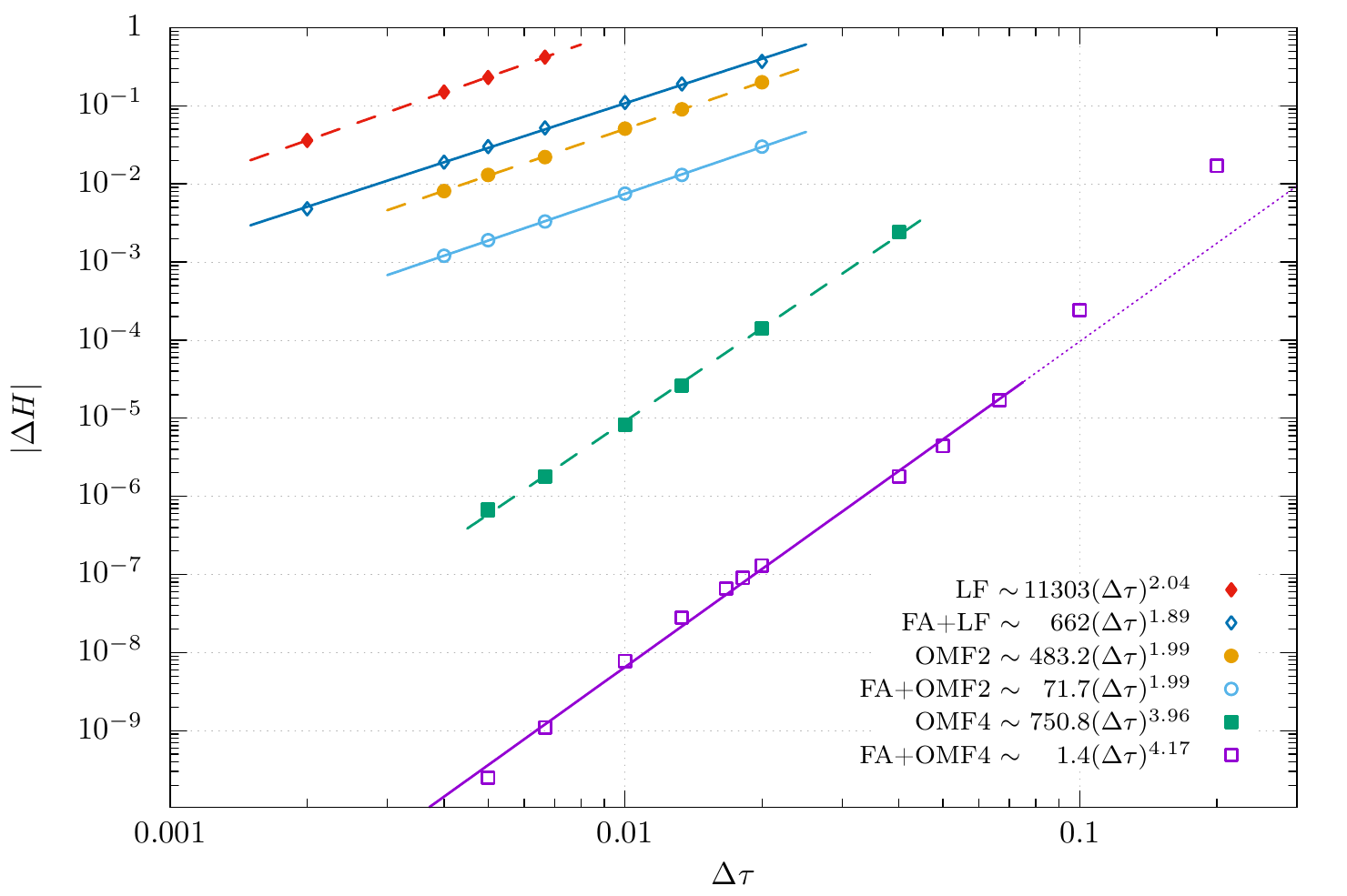}
   \caption{
   Violations of MD Hamiltonian conservation $\Delta H$ as a function of the MD
   integration step--size $\Delta \tau$, for all available integrators (LF,
   OMF2, OMF4), with and without Fourier Acceleration (FA). The lines represent
   the fit functions provided in the legend.
   }\label{fig:dH}
\end{figure} 

Figure~\ref{fig:dH} shows the violation $\Delta H$ as a function of $\Delta
\tau$ for all integrators, with and without Fourier Acceleration. A two
parameter function $\Delta H = a \, \Delta \tau^b$ has been fitted to the data
points. In all cases the obtained exponent is reasonably close to the expected
one. This test has been performed on a single thermalized configuration taken
from the \texttt{Q*D1} ensemble (table~\ref{tab:Q1}).

As expected there is a clear hierarchy among the three integrators. More
interestingly, Fourier Acceleration has the effect of reducing significantly
$\Delta H$. While no definite conclusion can be drawn from a single--configuration
experiment in this regard, the experience collected so far
suggests that this is quite generally the case: when Fourier Acceleration is
turned on, if one wants to keep the acceptance rate the same, larger values of
$\Delta \tau$ can be typically chosen. Obviously this does not mean that it is
always convenient to use Fourier Acceleration. In order to understand whether
this is the case, one should take into account the computational overhead and
the variation in autocorrelations. Fourier acceleration is known to reduce
significantly autocorrelations in the case of the free scalar theory, but also
in the case of non--compact pure U(1) theory~\cite{Borsanyi:2014jba}, which is a
theory of free photons. Nevertheless, as soon as a full QCD+QED simulation is
done, our experience suggests that autocorrelation times are quite insensitive
to the use of Fourier acceleration for the U(1) fields. These findings need
further investigations.

%% file: 46-deflation.tex
\begin{table}[tb]
  \small\centering
  \begin{tabular}{lllllllll}\toprule
             name &$\Nf$&$\beta$ & time bc  & $\VCs/2$         & $\kappa$ & $t_0/a^2$    & $a$\,[\fm] & $\mPS$\,[\MeV] \\\midrule
    \texttt{QCD1} & 2   & 5.3    & periodic & $64 \times 32^3$ & 0.136304 & ---          & 0.066 & 365  \\
    \texttt{Q*D1} & 2+1 & 3.55   & periodic & $32 \times 16^3$ & 0.137000 & $3.867(50)$  & 0.074 & 660  \\
  \bottomrule
  \end{tabular}
  \caption{%
           Details of test runs employing \Cstar boundary conditions in 3
           spatial directions.  Note that due to the \Cstar boundary
           conditions, the global (simulated) lattice \VCs is two times larger
           than the physical lattice because of the orbifold contruction.
           $\Nf=3$ simulations of \QCDQED (\texttt{Q*D1}) use the tree--level
           improved Symanzik gauge action (LW) for the \gSU gauge field with
           $\cswSU$ taken from~\cite{Bulava:2013cta}, and the Wilson plaquette
           action (W) for the electromagnetic field with $\cswU=1$.
           Furthermore, the electromagnetic coupling is set to $\alpem=0.05
           \approx 7 \alpem^\text{phys}$ with $\qel=1/6$, i.e., the doublet
           $(d\, s)_{-1/3}$ and $(u)_{+2/3}$ have been simulated.  The $\Nf=2$
           pure QCD simulation (\texttt{QCD1}) uses the plaquette action with
           non--perturbative $\cswSU$ of ref.~\cite{Jansen:1998mx}, and the
           lattice spacing was determined in ref.~\cite{Fritzsch:2012wq}. All
           runs have degenerate quarks with hopping parameter $\kappa$. Values
           for the neutral pseudoscalar mass $\mPS$ are given, as well as the
           flow time $t_0/a^2$ from which we naively derive the approximate
           lattice spacing of \texttt{Q*D1} using results of
           ref.~\cite{Bruno:2016plf}. 
          }\label{tab:Q1}
\end{table}
\begin{figure}[t]
 \centering
 {\includegraphics[width=0.9\textwidth,clip]{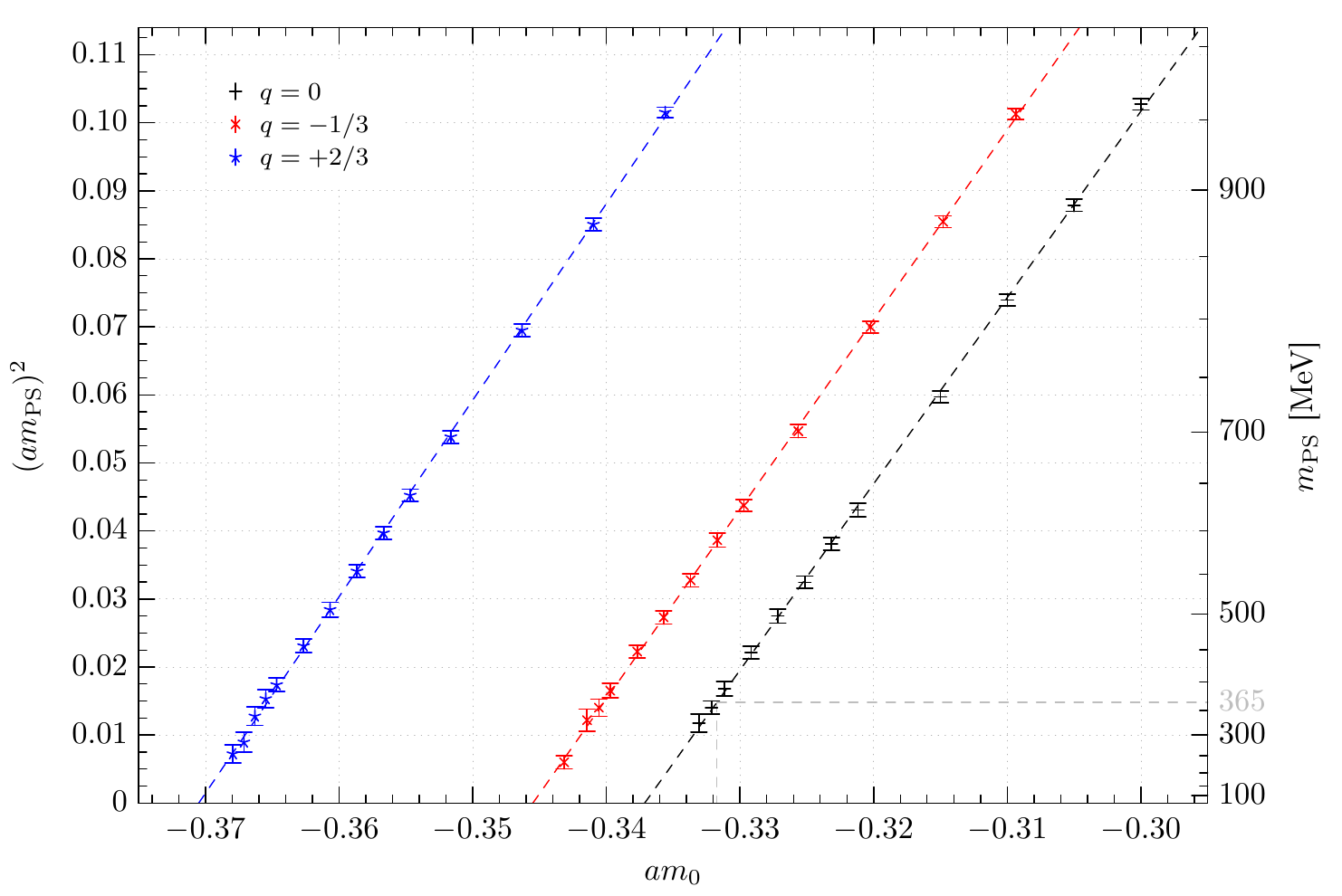}}
 \caption{%
         Mass of the $\bar{Q}' \gamma_5 Q$ valence pseudoscalar neutral meson
         has been calculated as a function of $q$ and $am_0=1/(2\kappa)-4$.
         QCD+qQED setup: SU(3) configurations are taken from the \texttt{QCD1}
         ensemble (table~\ref{tab:Q1}) and pure U(1) configurations are
         generated with $\alpem = 0.05$ and $q_\text{el}=1/6$. The dashed
         curves are fits to the expected (leading order) quark mass dependence,
         $[\mPS(q)]^2 = B(q) \{m_0-\mcr(q)\}$, and are shown only to
         guide the eye. The gray dashed line indicates the mass of the unitary
         point of the QCD simulation.
         }%
 \label{fig:mpi-vs-m0}
\end{figure}

The use of efficient solvers is a key factor in enabling simulations at quark
masses close to the physical point. The \oQxD code inherits all the solvers of
the \oQCD[-1.6] package: Conjugate Gradient (CG), Multi--Shift Conjugate
Gradient (MSCG), Generalized Conjugate Residual algorithm with
Schwartz--Alternating--Procedure as preconditioning (SAP+GCR), and a deflated
version of it (DFL+SAP+GCR). The deflated solver implements the idea of inexact
deflation introduced in~\cite{Luscher:2007se,Luscher:2007es} and an 
improvement involving inaccurate projection in the deflation preconditioner
proposed in~\cite{Frommer:2013fsa}.  

As the Dirac operator is passed as an argument to these solvers, their
implementation is blind to the coupling to the U(1) field and to \Cstar
boundary conditions. The efficiency of these solvers may be affected in
principle by the coupling to the U(1) field, i.e. may depend on the electric
charge of the Dirac operator. However this turns out not to be the case. The
goal of this section is to describe two tests in support of this statement.
These tests have been run on \textit{Altamira} HPC at IFCA-CSIC, which consists
of 158 computing nodes, each of them with two
Intel\textsuperscript{\textregistered} Xeon processors (E5-2670) at 2.6\,GHz.
Nodes are connected with Mellanox\textsuperscript{\textregistered} Infiniband
FDR (56\,Gb/s). 

An electroquenched (QCD+qQED) setup has been considered for both tests, with
SU(3) configurations from the \texttt{QCD1} ensemble (table~\ref{tab:Q1}) and
pure U(1) configurations generated with $\alpem = 0.05$ and $\qel=1/6$. Two
degenerate valence quarks $Q$ and $Q'$ have been considered, with electric
charge $q$ and bare mass $m_0$. The mass $\mPS$ of the $\bar{Q}' \gamma_5 Q$
valence pseudoscalar neutral meson has been calculated as a function of $q$ and
$m_0$ and is shown in fig.~\ref{fig:mpi-vs-m0}. Notice that the critical bare
mass depends very heavily on the electric charge, as expected. For this reason
it makes sense to compare the solver performance for different electric charges
keeping fixed the value of $\mPS$ (rather than the bare mass).

\begin{figure}[t]
 \centering
 {\includegraphics[width=0.85\textwidth,clip]{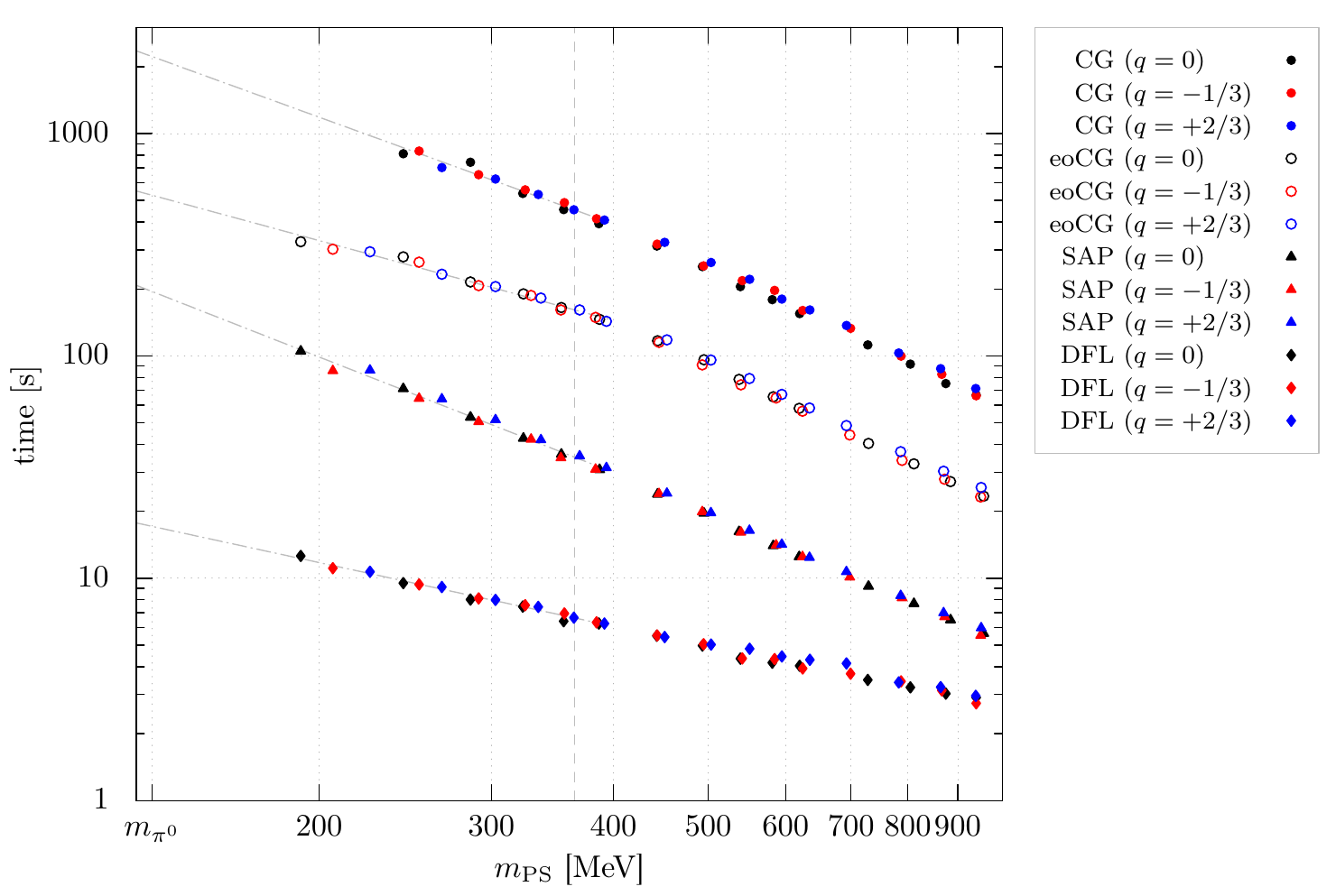}}
 \caption{%
        Comparison of performance of various solvers and various electric
        charges as a function of the mass $\mPS$ of the valence neutral pion.
        In all cases, the inverse of the even--odd preconditioned Dirac operator
        has been calculated on random sources. One representative QCD+qQED
        configuration has been used (SU(3) configuration from the \texttt{QCD1}
        ensemble, table~\ref{tab:Q1}, and pure U(1) configuration generated
        with $\alpem = 0.05$ and $\qel=1/6$). The same residue of $10^{-10}$
        has been chosen for the three solvers. The solver performance is
        insensitive to the electric charge.
        }%
 \label{fig:dflvsCG}
\end{figure}

In the first test, the time needed to invert the even--odd preconditioned Dirac
operator (with a representative QCD+qQED configuration) on 15 random sources has
been measured, using the CG, SAP+GCR, and DFL+SAP+GCR solvers. The shortest time
has been plotted in fig.~\ref{fig:dflvsCG} for electric charges $q=0,-1/3,2/3$
and a range of values of $\mPS$. It is evident that the performance of
all solvers is insensitive to the electric charge.

One important caveat needs to be pointed out for the DFL+SAP+GCR solver. Before
applying this solver, one needs to generate the deflation subspace, which is
constructed from approximate eigenvectors of the Dirac operator. The code allows
the possibility to choose different parameters for the Dirac operator used in
the solver and the one used to generate the deflation subspace. This is very
useful in practice since having a slightly heavier bare mass or even a twisted
mass for the generation of the deflation subspace generally speeds up the
calculation without affecting the performance of the solver. On the other hand,
it is crucial to generate the deflation subspace with the same electric charge
of the Dirac operator that needs to be inverted. If this is not done, the
DFL+SAP+GCR solver loses efficiency dramatically. For this reason, in contrast
to \oQCD[-1.6], the \oQxD code can handle simultaneously
several deflation subspaces. These deflation subspaces can be generated with
different parameters and will all be updated during the MD evolution. The user
can specify in the input file which deflation subspace should be used for each
DFL+SAP+GCR solver independently. In practice, in a realistic QCD+QED
simulation, one would need to generate only two deflation subspaces, one for
up--type quarks and one for down--type quarks. It has been checked also that the
time needed to generate the deflation subspace is insensitive to the electric
charge as long as $\mPS$ is kept fixed.

\begin{figure}[t]
 \centering
 {\includegraphics[width=0.85\textwidth,clip]{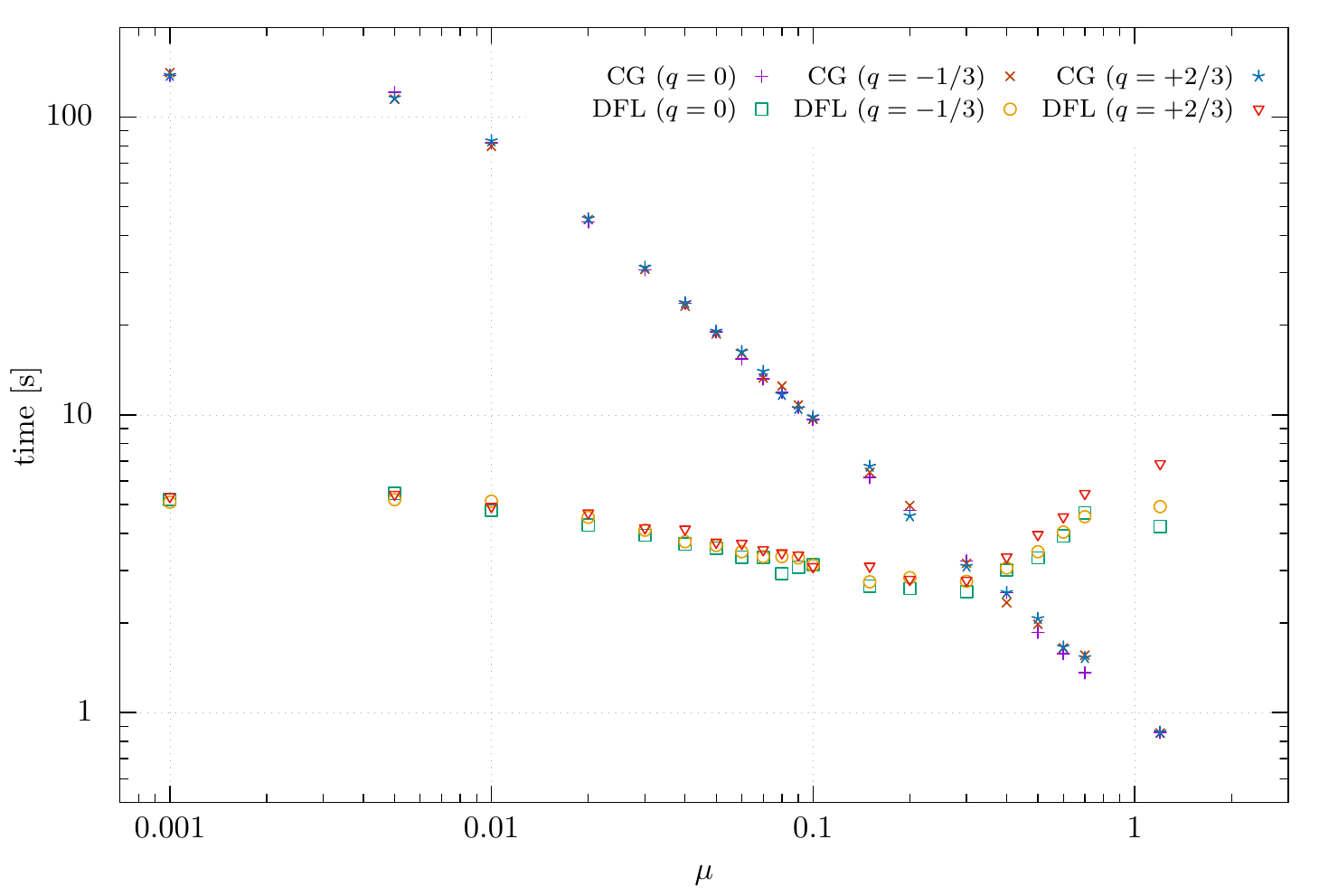}}
 \caption{%
        Comparison of performance of various solvers and various electric
        charges as a function of the twisted mass $\mu$. In all cases, the
        inverse of $(\Dhat^\dag \Dhat + \mu^2)$ has been calculated on random
        sources. The mass of the valence neutral pion (calculated at $\mu=0$)
        has been chosen to be $\mPS \simeq 329\,\MeV$. One representative
        QCD+qQED configuration has been used (SU(3) configuration from the
        \texttt{QCD1} ensemble, table~\ref{tab:Q1}, and pure U(1) configuration
        generated with $\alpem = 0.05$ and $\qel=1/6$). The same residue of
        $10^{-8}$ has been chosen for the three solvers. The solver performance
        is insensitive to the electric charge. As expected, the deflated solver
        loses efficiency at large values of $\mu$ and eventually fails to
        converge for the three highest value.
        }%
 \label{fig:dfl}
\end{figure}

In the second test, a single value of $\mPS \simeq 354\,\MeV$ has been chosen,
and the time needed to invert $(\Dhat^\dag \Dhat + \mu^2)$ has been measured for
various values of the twisted mass $\mu$, using the CG and DFL+SAP+GCR solvers.
One representative QCD+qQED configuration and 48 random sources have been used.
The shortest time has been plotted in fig.~\ref{fig:dfl} for electric charges
$q=0,-1/3,2/3$ and a range of values of $\mu$. The inversion of $(\Dhat^\dag
\Dhat + \mu^2)$ is relevant to calulate the rational approximation of
non--integer powers of $\Dhat^\dag \Dhat$ (see section~\ref{sec:tech}). Also in
this case, the performance of the two solvers is seen to be insensitive to the
electric charge as long as $\mPS$ is kept fixed.

%% file: 48-observables.tex
\begin{figure}[p!]
 \centering
 {\includegraphics[width=0.495\textwidth,clip,page=1]{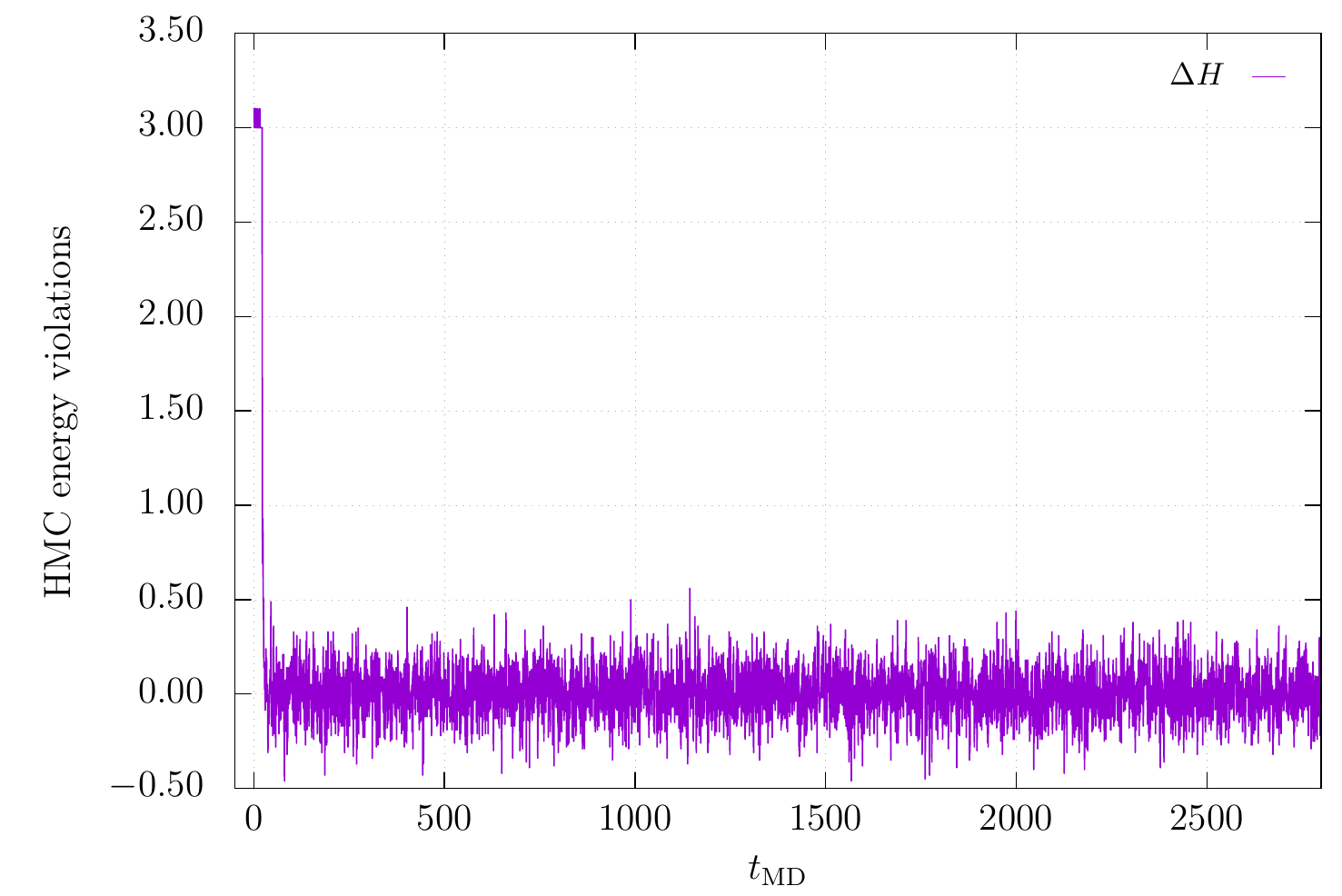}}\hfill
 {\includegraphics[width=0.495\textwidth,clip,page=2]{figures/obs_plots.pdf}}\\
 {\includegraphics[width=0.495\textwidth,clip,page=3]{figures/obs_plots.pdf}}\hfill
 {\includegraphics[width=0.495\textwidth,clip,page=4]{figures/obs_plots.pdf}}
 {\includegraphics[width=0.495\textwidth,clip,page=5]{figures/obs_plots.pdf}}\hfill
 {\includegraphics[width=0.495\textwidth,clip,page=6]{figures/obs_plots.pdf}}\\
 {\includegraphics[width=0.495\textwidth,clip,page=7]{figures/obs_plots.pdf}}\hfill
 {\includegraphics[width=0.495\textwidth,clip,page=8]{figures/obs_plots.pdf}}
 \caption{%
         Selected observables for simulation \texttt{Q*D1} including
         thermalisation part. Left--right/top--bottom: HMC energy violations
         $\Delta H$, average plaquette for SU(3) and U(1) gauge fields, energy
         density $E(t)$ for SU(3), energy density for U(1), topological charge
         $Q(t)$, lowest eigenvalue $\hat\lambda_{\rm min}$ in the spectrum of
         $|\gamma_5 \Dhat|$, and reweighting factors $W_q$ for two different
         numerical accuracies, $\delta=O(10^{-11})$ (left) and
         $\delta=O(10^{-9})$ (right).
         }\label{fig:obs}%
\end{figure}

Beside the electroquenched tests in the previous section, a new set of tests
is done using dynamical QCD+QED simulations with Wilson fermions and
\Cstar~boundary conditions. The dynamical degrees of freedom of the U(1) gauge
field are included in the simulation labeled \texttt{Q*D1} in
table~\ref{tab:Q1}. \texttt{Q*D1} takes over the parameters from the
H200 ensemble of the $\Nf=2+1$ CLS~\cite{Bruno:2014jqa} effort, except that the
lattice extent is halved in each of the space--time directions.  As the
dynamical U(1) degrees of freedom contribute to the renormalization of the bare
parameters, the estimate for the lattice spacing and pion mass cannot be taken
over from the CLS ensembles,%
\footnote{%
Had the U(1) d.o.f. been switched off ($\alpem = 0$), the chosen parameter set
would correspond to $a\approx 0.064\,\fm$ and $\mPS \approx 420\,\MeV$.
}
but rather need to be estimated independently. Such an endeavour is beyond the
scope of this paper. However, an estimate for $t_0/a^2$ is given in
table~\ref{tab:Q1} for future reference. The reference flow time $t_0$ is
implicitly given by $[t_0^2\langle E(t_0)\rangle]=0.3$ using the Wilson flow
and clover discretisation of the SU(3) field strength tensor in the definition
of the energy density $E(t)$~\cite{Luscher:2010iy}. A rough estimate of $a$ is
given after naively matching $t_0/a^2$ to the data provided in table~III of
ref.\cite{Bruno:2016plf}. 

Although \oQxD allows for twisted--mass reweighting, that option is not
required for \texttt{Q*D1} (${\mu}=0.0$).  All three bare sea quark masses,
$am_{0,i}=1/(2\kappa_i)-4$, are taken to be degenerate. As demonstrated in the
previous section and shown in fig.~\ref{fig:mpi-vs-m0}, this necessarily leads
to a large difference in the neutral pseudoscalar masses due to the differences
in quark charges. One thus ends up with a degenerate pair of down--type quarks
($q=-1/3$), and a single but significantly heavier up--type quark ($q=2/3$).
Hence, the simulations are essentially probing a somewhat unphysical version of
the $\Nf=2+1$ theory, but are sufficient to probe standard observables and
performance of the code.

In fig.~\ref{fig:obs} a summary of selected observables is given for simulation
\texttt{Q*D1}. 
The run was stable and did not show any particular issue during the course of
the simulation.  Most of the observables presented in the following include the
thermalisation part.  Starting from a random configuration, the HMC energy
violations, measured every trajectory ($\tau=0.7$ MDU), drop after a few
iterations and stably fluctuate in the range $[-0.5,+0.5]$. The simulation
employs the OMF4 integrator without Fourier acceleration and the spectral
ranges of the individual quark flavours have been properly set.  Next the
average plaquette for the SU(3) and U(1) gauge fields are presented.  The
former is shifted by a constant amount for better comparison. The SU(3)
plaquette has much larger statistical fluctuations and requires longer
thermalisation times than the U(1) plaquette even without Fourier acceleration.
The next two plots show the two available definitions of the (renormalized)
energy density $E(t)$ at a flow time $t=3.2$ for the SU(3) and U(1) part,
respectively. 
The topological charge $Q$ (measured at the same flow time) fluctuates well
after rapid changes during the thermalisation phase of the run. The smallest
eigenvalues of $|\gamma_5 \Dhat_\mathrm{u}|$ and $|\gamma_5
\Dhat_\mathrm{d/s}|$ follow, confirming that the lower end of the spectral
ranges of the rational approximations have been chosen correctly. No
exceptionally small values are present, which is not surprising considering the
heavy pseudoscalar mass simulated here. 

The \texttt{Q*D1} run has been produced with a rational approximation with
relative precision $\delta = O(10^{-11})$. A second run has been performed with
the same parameters as \texttt{Q*D1} except for the rational approximation,
which has been chosen with relative precision $\delta = O(10^{-9})$. The
logarithms of the reweighing factors for both runs are shown in the last two
panes of fig.~\ref{fig:obs}. As expected, the reweighting factor for the run
with a better rational appriximation is closer to 1 (and its logarithm is
closer to 0).

%% file: 90-summary.tex
We presented \oQxD~\cite{openQxD-csic}, the first open source package which
allows to perform full lattice simulations of QCD$+$QED, QCD or QED. The code
implements the proposal of ref.~\cite{Lucini:2015hfa} and allows to choose
\Cstar boundary conditions along the spatial directions but also periodic
boundary conditions can be simulated efficiently. Moreover, the chosen theory
can be simulated by choosing either periodic, Schr\"odinger Functional or open
boundary conditions along the time direction.

The new code is based on the \oQCD~\cite{openQCD} package from which it
inherits the highly optimized implementation of the Dirac operator, of the
solvers, of the HMC and of the RHMC algorithms. The \oQxD package extends the
algorithmic functionalities of the \oQCD code by giving the possibility of
using multiple deflation subspaces in a single simulation, of implementing
rational approximations of generic powers of the Dirac operator (with and
without twisted--mass preconditioning) and by implementing Fourier Acceleration
for the evolution of the U(1) field. 

We presented the main functionalities of the code and discussed the theoretical
motivations behind the algorithmic choices and their specific implementations.
We also presented a guide to instruct the user to run a full QCD$+$QED
simulation with \oQxD and discussed the results of some tests. These include
low--level tests aiming at assessing the correctness of the implementation of
the different algorithms but also some benchmarks to measure the performance of
the code. 

Given the good performance and high scalability on modern supercomputing
cluster architectures, \oQxD can profitably be used to generate QCD$+$QED gauge configurations with \Cstar boundary conditions (but not only) in a realistic setup with the aim
of computing QED radiative corrections to phenomenologically relevant
observables.

%% file: acknow.tex
%
%
The simulations were performed on the following HPC systems: Altamira, provided
by IFCA at the University of Cantabria; FinisTerrae~II, provided by CESGA
(Galicia Supercomputing Centre); the Lonsdale cluster maintained by the Trinity
Centre for High Performance Computing; and the Lattice-HPC cluster at CERN.
FinisTerrae~II was funded by the Xunta de Galicia and the Spanish MINECO under
the 2007--2013 Spanish ERDF. Lonsdale was funded through grants from the
Science Foundation Ireland. We thankfully acknowledge the computer resources
offered and the technical support provided by the staff of these computing
centers. 
We thank the Theoretical Physics Department at CERN for
hospitality during the workshop \emph{Advances in Lattice Gauge Theory 2019}, 
allowing us to jointly finalise the present work.

%% file: 98-rhmc.tex
\subsection{Rational approximation}\label{sec:ratapprox}

It is convenient to introduce the hermitian operator $\hat{Q}=\gamma_5 \hat{D}$,
in terms of which $\hat{D}^\dag \hat{D} = \hat{Q}^2$. Assume that the spectrum
of $|\hat{Q}|$ is contained in the interval $[r_a,r_b]$, and choose an integer
$n$. A rational function of order $[n,n]$ in $q^2$ has the form
\begin{gather}
   \rho(q^2) = A \prod_{j=1}^n \frac{q^2 + \nu_j^2}{q^2 + \mu_j^2} \ .
   \label{eq:ratfunc}
\end{gather}
Without loss of generality one can assume
\begin{gather}
   \nu_1 > \nu_2 > \dots > \nu_n \ , \qquad
   \mu_1 > \mu_2 > \dots > \mu_n \ .
\end{gather}

$\rho(q^2)$ is chosen to be the optimal rational approximation of order $[n,n]$
of the function $(q^2 + \hat{\mu}^2)^{-\alpha}$ in the domain $q \in [r_a,r_b]$,
i.e. the rational function of the form~\eqref{eq:ratfunc} which minimizes the
uniform relative error
\begin{gather}
   \delta = \max_{q \in [r_a,r_b]} | 1 - (q^2 + \hat{\mu}^2)^{\alpha} \rho(q^2) |
   \ .
   \label{eq:error}
\end{gather}
As explained in sec. \ref{sec:input}, the optimal rational approximation can be
calculated with the \texttt{minmax} code which implements the minmax
approximation algorithm in multiple precision.

If $\rho(q^2)$ is the desired optimal rational approximation, the operator
$\Rmu$ which appears in eq.~\eqref{eq:simdistr} is defined simply as
\begin{gather}
   \Rmu = \rho(\hat{Q}^2) = \rho(\hat{D}^\dag \hat{D}) =
   A \prod_{j=1}^n \frac{\hat{D}^\dag \hat{D} + \nu_j^2}{\hat{D}^\dag \hat{D} + \mu_j^2}
   \ .
\end{gather}
Eq.~\eqref{eq:error} implies the following norm bound
\begin{gather}
   \| 1 - (\hat{D}^\dag \hat{D} + \hat{\mu}^2)^{\alpha} \Rmu \| \le \delta \ .
   \label{eq:normbound}
\end{gather}

\subsection{Frequency splitting and pseudofermion action}\label{subsec:freq}

\oQxD inherits from \oQCD the frequency splitting of the rational approximation:
the factors of the rational approximation can be split in different
pseudofermion actions; the corresponding forces can be included in different
levels of the MD integrator, providing a useful handle to optimize the
algorithm. This procedure is similar to the Hasenbusch decomposition for the HMC
algorithm~\cite{Hasenbusch:2001ne}.

The rational approximation constructed in section~\ref{sec:ratapprox} is broken
up in factors of the form
\begin{gather}
   P_{k,l} = \prod_{j=k}^l \frac{\hat{D}^\dag \hat{D} + \nu_j^2}{\hat{D}^\dag \hat{D} + \mu_j^2} \ .
   \label{eq:Pfactor}
\end{gather}
For example, if $n = 12$ a possible factorization is
\begin{gather}
   \Rmu = A P_{1,5} P_{6,9} P_{10,12} \ .
\end{gather}
The contribution of $\Rmu$ to the quark determinant is
\begin{gather}
   \det \Rmu^{-1} = \text{constant} \times \det P_{1,5}^{-1} \ \det P_{6,9}^{-1} \ \det P_{10,12}^{-1} \ .
\end{gather}
Each $P_{k,l}^{-1}$ determinant is simulated as usual by adding a pseudofermion
action of the form
\begin{gather}
   S_{\text{pf},k,l} = ( \phi^{k,l}_\text{e} , P_{k,l} \phi^{k,l}_\text{e} ) \ ,
   \label{eq:Spf}
\end{gather}
where the fields $\phi^{k,l}_\text{e}$ are independent pseudofermions that live
on the even sites of the lattice. By using a partial fraction decomposition
\begin{align}
   P_{k,l}  &=  1 + \sum_{j=k}^l \frac{\sigma_j}{\hat{D}^\dag \hat{D} + \mu_j^2}  \;, \\
   \sigma_j &= (\nu_j^2 - \mu_j^2) \prod_{\substack{m=l,\dots,k\\m \neq j}} 
               \frac{\nu_m^2-\mu_j^2}{\mu_m^2-\mu_j^2}  \;,
\end{align}
the pseudofermion action in eq.~\eqref{eq:Spf} is cast into a sum of terms of
the type
\begin{gather}
   S_{\text{pf},k,l} = ( \phi^{k,l}_\text{e} , \phi^{k,l}_\text{e} ) 
                        + \sum_{j=k}^l \sigma_j \ ( \phi^{k,l}_\text{e}, (\hat{D}^\dag \hat{D} + \mu_j^2)^{-1} \phi^{k,l}_\text{e} ) \ .
   \label{eq:Spf2}
\end{gather}

\subsection{Reweighting factors}\label{sec:app-reweighting}

Let $\R$ and $\Rmu$ be the optimal rational approximations of order $[n,n]$ for
$(\hat{D}^\dag \hat{D})^{-\alpha}$ and $(\hat{D}^\dag \hat{D} +
\hat{\mu}^2)^{-\alpha}$ respectively. It is assumed that the relative errors of
the two rational approximations are not greater than $\delta$ in the common
spectral range $[r_a,r_b]$.

The reweighting factor $W$ defined in eq.~\eqref{eq:W} is decomposed in two
factors which are calculated separately, i.e.
\begin{align}
   W            &= W_\text{rat} W_\text{rtm}                   \;, \\
   W_\text{rat} &= \det [ (\hat{D}^\dag \hat{D})^{\alpha} \R ] \;, \label{eq:Wrat} \\
   W_\text{rtm} &= \det [ \R^{-1} \Rmu ]                       \;. \label{eq:Wtm}
\end{align}

\subsubsection{Reweighting factor $W_\text{rat}$}\label{sec:tmrew}

In the calculation of the reweighting factor $W_\text{rat}$ in
eq.~\eqref{eq:Wrat}, it is assumed that the exponent $\alpha$ is a positive
rational number of the form
\begin{gather}
   \alpha = \frac{u}{v} \ ,
\end{gather}
where $u$ and $v$ are natural numbers. The reweighting factor can be
represented as
\begin{gather}
   W_\text{rat} = \det [ \hat{Q}^{2u} \R^v ]^{\frac{1}{v}}
   =
   \det (1 + Z)^{\frac{1}{v}} \ ,
   \label{eq:Wrat2}
\end{gather}
where the operator $Z$ is defined as
\begin{gather}
   Z = \hat{Q}^{2u} \R^v - 1 \ .
\end{gather}
The determinant in eq.~\eqref{eq:Wrat2} is estimated stochastically
\begin{gather}
   W_\text{rat} = \lim_{N \to \infty} \frac{1}{N} \sum_{j=1}^N \exp \{ - ( \eta^j_\text{e} , [ (1 + Z)^{-\frac{1}{v}} - 1 ] \eta^j_\text{e} ) \}
   \ ,
   \label{eq:Wrat3}
\end{gather}
where the fields $\eta^j_\text{e}$ are $N$ independent normally--distributed
pseudofermions that live on the even sites of the lattice. From the norm bound
in eq.~\eqref{eq:normbound} for $\hat{\mu}=0$, and the positivity of $\R$
(which is guaranteed if the relative error $\delta$ is small enough), it
follows that
\begin{gather}
   0 \le 1+Z = \hat{Q}^{2u} \R^v = [ \hat{Q}^{2\alpha} \R]^v \le (1 + \delta)^v \ ,
\end{gather}
which yields the norm bound
\begin{gather}
   \| Z \| \le \Delta = (1 + \delta)^v - 1 = v \delta + O(\delta^2) \ .
   \label{eq:normbound2}
\end{gather}
Therefore the Taylor series
\begin{align}
   (1 + Z)^{-\frac{1}{v}} &= 1 + \sum_{n=1}^\infty c_{v,n} \, Z^n \;, &
                  c_{v,n} &= (-1)^n \frac{\tfrac{1}{v} (\tfrac{1}{v}+1) \cdots (\tfrac{1}{v} + n - 1)}{n!} \;,
\end{align}
converges rapidly in operator norm. The exponent in eq.~\eqref{eq:Wrat3} can be
estimated from the first few terms of
\begin{gather}
   ( \eta^j_\text{e} , [ (1 + Z)^{-\frac{1}{v}} - 1 ] \eta^j_\text{e} ) = \sum_{n=1}^\infty
   c_{v,n} \, ( \eta^j_\text{e} , Z^n \eta^j_\text{e} )  \ .
\end{gather}
It is possible to estimate the size of these terms by noting that $\|
\eta^j_\text{e} \|^2$ is very nearly equal to 12 times the number $N_\text{e}$
of even lattice points. Taking the bound~\eqref{eq:normbound2} into account,
the following estimate is obtained
\begin{gather}
   | ( \eta^j_\text{e} , Z^n \eta^j_\text{e} ) | \le \| Z \|^n \, \| \eta^j_\text{e} \|^2
   \le
   \Delta^n \| \eta^j_\text{e} \|^2
   \simeq
   12 (v \delta)^n N_\text{e}
   \ .
\end{gather}

The statistical fluctuations of the exponents in eq.~\eqref{eq:Wrat3} derive
from those of the gauge field and those of the random sources
$\eta^j_\text{e}$. For a given gauge field, the variance of the exponent is
equal to
\begin{gather}
   \tr \{ [ (1 + Z)^{-\frac{1}{v}} - 1 ]^2 \}
   =
   \frac{1}{v^2} \tr Z^2 + O(\delta^3)
   \le
   12 N_\text{e} \delta^2 + O(\delta^3) \ .
\end{gather}
These fluctuations are guaranteed to be small if, for instance, $12 N_\text{e}
\delta^2 \le 10^{-4}$. One can then just as well set $N = 1$ in
eq.~\eqref{eq:Wrat3}, i.e. a sufficiently accurate stochastic estimate of
$W_\text{rat}$ is obtained in this case with a single random source.

When the stronger constraint $12 N_\text{e} \delta \le 10^{-2}$ is satisfied,
the reweighting factor $W_\text{rat}$ deviates from 1 by at most 1\%. Larger
approximation errors can however be tolerated in practice as long as the
fluctuations of $W_\text{rat}$ remain small.

\subsubsection{Reweighting factor $W_\text{rtm}$}

Let us choose a rational approximation $\Rmu$ of order $[n,n]$ for
$(\hat{D}^\dag \hat{D} + \hat{\mu}^2)^{-\alpha}$ of the form
\begin{gather}
   \Rmu =
   A \prod_{j=1}^n \frac{\hat{D}^\dag \hat{D} + \nu_j^2}{\hat{D}^\dag \hat{D} + \mu_j^2}
   \ , \\
   \nu_1 > \nu_2 > \dots > \nu_n \ , \qquad
   \mu_1 > \mu_2 > \dots > \mu_n \ ,
\end{gather}
and a rational approximation $\R$ of order $[n,n]$ for $(\hat{D}^\dag
\hat{D})^{-\alpha}$ of the form
\begin{gather}
   \R =
   \tilde{A} \prod_{j=1}^n \frac{\hat{D}^\dag \hat{D} + \tilde{\nu}_j^2}{\hat{D}^\dag \hat{D} + \tilde{\mu}_j^2}
   \ , \\
   \tilde{\nu}_1 > \tilde{\nu}_2 > \dots > \tilde{\nu}_n \ , \qquad
   \tilde{\mu}_1 > \tilde{\mu}_2 > \dots > \tilde{\mu}_n \ .
\end{gather}
Let us rewrite eq.~\eqref{eq:Wtm} as
\begin{gather}
   W_\text{rtm} = \det [ \Rmu^{-1} \R ]^{-1} \ .
\end{gather}
Notice that the operator $\Rmu^{-1} \R$ is also a rational function of
$\hat{Q}^2=\hat{D}^\dag \hat{D}$. It is convenient to break up this rational
function in factors of the type
\begin{gather}
   \tilde{P}_{k,l} = \prod_{j=k}^l \frac{(\hat{D}^\dag \hat{D} + \mu_j^2)(\hat{D}^\dag \hat{D} + \tilde{\nu}_j^2)}{(\hat{D}^\dag \hat{D} + \nu_j^2)(\hat{D}^\dag \hat{D} + \tilde{\mu}_j^2)} \ .
\end{gather}
If $n = 12$, for example, the reweighting factor $W_\text{rtm}$ can be factorized as
\begin{gather}
   W_\text{rtm} = \text{constant} \times \det \tilde{P}_{1,5}^{-1} \ \det \tilde{P}_{6,9}^{-1} \ \det \tilde{P}_{10,12}^{-1} \ .
\end{gather}
Each of the above determinants is estimated stochastically
\begin{gather}
   \det \tilde{P}_{k,l}^{-1} = \lim_{N \to \infty} \frac{1}{N} \sum_{j=1}^N \exp \{ - ( \eta^j_\text{e} , [ \tilde{P}_{k,l} - 1 ] \eta^j_\text{e} ) \}
   \ ,
\end{gather}
where the fields $\eta^j_\text{e}$ are $N$ independent normally--distributed
pseudofermions that live on the even sites of the lattice. It is useful to
consider the partial fraction decomposition
\begin{align}
   \tilde{P}_{k,l}  &= 1 + \sum_{j=k}^l \left( \frac{\sigma_j}{\hat{D}^\dag \hat{D} + \nu_j^2} 
                         + \frac{\tilde{\sigma}_j}{\hat{D}^\dag \hat{D} + \tilde{\mu}_j^2} \right)   \;, \\
   \sigma_j         &= \frac{(\tilde{\nu}_j^2-\nu_j^2)(\mu_j^2-\nu_j^2)}{\tilde{\mu}_j^2-\nu_j^2}   
                       \prod_{\substack{m=l,\dots,k\\m \neq j}} \frac{(\tilde{\nu}_m^2-\nu_j^2)(\mu_m^2-\nu_j^2)}{(\tilde{\mu}_m^2-\nu_j^2)(\nu_m^2-\nu_j^2)}   \;, \\
   \tilde{\sigma}_j &= \frac{(\tilde{\nu}_j^2-\tilde{\mu}_j^2)(\mu_j^2-\tilde{\mu}_j^2)}{\nu_j^2-\tilde{\mu}_j^2}
                       \prod_{\substack{m=l,\dots,k\\m \neq j}} \frac{(\tilde{\nu}_m^2-\tilde{\mu}_j^2)(\mu_m^2-\tilde{\mu}_j^2)}{(\tilde{\mu}_m^2-\tilde{\mu}_j^2)(\nu_m^2-\tilde{\mu}_j^2)}   \;.
\end{align}
Typically $\sigma_j$ and $\tilde{\sigma}_j$ are found to have opposite signs.
Also, for small values of $j$, $|\sigma_j|$ and $|\tilde{\sigma}_j|$ are of the
same order of magnitude, therefore it is convenient for numerical stability to
use the following representation
\begin{gather}
   \tilde{P}_{k,l}
   =
   1 + \sum_{j=k}^l \frac{(\sigma_j+\tilde{\sigma}_j) (\hat{D}^\dag \hat{D}) + \sigma_j \tilde{\mu}_j^2+\tilde{\sigma}_j \nu_j^2}{(\hat{D}^\dag \hat{D} + \nu_j^2)(\hat{D}^\dag \hat{D} + \tilde{\mu}_j^2)}
   \ .
\end{gather}

%% file: 99-fourier.tex
The U(1) momentum is generally represented in momentum space as
\begin{gather}
   \pi(x,\mu) = \frac{1}{L^3} \sum_{p_0 \in E_\mu} \sum_{\vec{p} \in \mathcal{P}} e_\mu(p_0,x_0) e^{i \vec{p} \vec{x}} \tilde{\pi}(p,\mu) \ .
\end{gather}
The basis functions $e_\mu(p_0,x_0)$ (for fixed $\mu$) are orthogonal with
respect to a weighted scalar product
\begin{gather}
   \sum_{x_0} w_\mu(x_0) e_\mu^*(p_0,x_0) e_\mu(q_0,x_0) = \delta_{p_0,q_0} \ ,
\end{gather}
where the weight $w_\mu(x)$ is taken to be $1/2$ if $x$ belongs to an open
boundary (i.e. $x_0=0$ for open and open--SF b.c.s, and $x_0=T-1$ for open
b.c.s) and $\mu=1,2,3$. In all other cases $w_\mu(x)$ is taken to be $1$. The
relation between $\pi$ and $\tilde{\pi}$ is easily inverted
\begin{gather}
   \tilde{\pi}(p,\mu)
   =
   \sum_{x} w_\mu(x_0) e_\mu^*(p_0,x_0) e^{-i \vec{p} \vec{x}} \pi(x,\mu) \ .
\end{gather}
The set $\mathcal{P}$ is given by all spatial momenta $\vec{p} = (p_1,p_2,p_3)$
of the form
\begin{gather}
   p_k = \tfrac{\pi}{L_k} (2n_k+c_k) \quad \text{with } n_k=0,\dots,L_k-1 \ ,
\end{gather}
where $c_k=0$ if $k$ is a periodic direction and $c_k=1$ if $k$ is a \Cstar
direction. The sets $E_\mu$ and the eigenfunctions $e_\mu(p_0,x_0)$ depend on
the boundary conditions in time. In the following $k=1,2,3$.
\begin{itemize}
   \item Open boundary conditions:
   \begin{gather}
      E_0 = \frac{\pi}{N_0-1} \{ 1 , \dots , N_0-1 \} \ , \qquad
      E_k = \frac{\pi}{N_0-1} \{ 0 , \dots , N_0-1 \} \ , \\
      e_0(p_0,x_0) = \frac{i}{(1+\delta_{p_0,\pi})(N_0-1)} \sin \left[ p_0\left(x_0 + \frac{1}{2} \right) \right] \ , \\
      e_k(p_0,x_0) = \frac{1}{(1+\delta_{p_0,0}+\delta_{p_0,\pi})(N_0-1)} \cos ( p_0 x_0 ) \ .
   \end{gather}
   \item SF boundary conditions:
   \begin{gather}
      E_0 = \frac{\pi}{N_0} \{ 0 , \dots , N_0-1 \} \ , \qquad
      E_k = \frac{\pi}{N_0} \{ 1 , \dots , N_0-1 \} \ , \\
      e_0(p_0,x_0) = \frac{1}{(1+\delta_{p_0,\pi})N_0} \cos \left[ p_0\left(x_0 + \frac{1}{2} \right) \right] \ , \\
      e_k(p_0,x_0) = \frac{i}{N_0} \sin ( p_0 x_0 ) \ .
   \end{gather}
   \item Open-SF boundary conditions:
   \begin{gather}
      E_0 = E_k = \frac{\pi}{N_0} \left( \{ 0 , \dots , N_0-1 \} + \frac{1}{2} \right) \ , \\
      e_0(p_0,x_0) = \frac{i}{N_0} \sin \left[ p_0\left(x_0 + \frac{1}{2} \right) \right] \ , \\
      e_k(p_0,x_0) = \frac{1}{N_0} \cos \left[ p_0\left(x_0 + \frac{1}{2} \right) \right] \ .
   \end{gather}
   \item Periodic boundary conditions:
   \begin{gather}
      E_0 = E_k = \frac{2\pi}{N_0} \{ 0 , \dots , N_0-1 \} \ , \\
      e_0(p_0,x_0) = e_k(p_0,x_0) = \frac{1}{N_0} \exp ( i p_0 x_0 ) \ .
   \end{gather}
\end{itemize}
We use the Fourier decomposition to define the intermediate operator
$D_\text{N}$
\begin{gather}
   [D_\text{N}\pi](x,\mu) = \frac{1}{L^3} \sum_{p_0 \in E_\mu} \sum_{\vec{p} \in \mathcal{P}} e_\mu(p_0,x_0) e^{i \vec{p} \vec{x}} \tilde{D}_\text{N}(p) \tilde{\pi}(p,\mu) \ , \\
   \tilde{D}_\text{N}(p) =
   \begin{cases}
      1 & \text{if } p=0 \\
      4 \sum_\mu \sin^2 \frac{p_\mu}{2} \quad & \text{otherwise}
   \end{cases}
   \ .
\end{gather}
Explicity
\begin{gather}
   D_\text{N}(x,\mu;y,\nu)
   =
   \frac{1}{L^3} \sum_{p_0 \in E_\mu} \sum_{\vec{p} \in \mathcal{P}}
   \tilde{D}_\text{N}(p) \delta_{\mu\nu}
   e^{i \vec{p} (\vec{x}-\vec{y})}
   e_\mu(p_0,x_0) e_\nu^*(p_0,y_0) w_\nu(y_0)
   \ .
\end{gather}
With respect to the scalar product defined by
\begin{gather}
   ( \phi , \phi )_G = (\phi , G \phi) \ , \\
   [G\phi](x,\mu) = w_\mu(x_0) \phi(x,\mu) \ .
\end{gather}
the operator $D_\text{N}$ is symmetric and strictly positive, i.e.
\begin{gather}
   ( \phi' , D_\text{N} \phi )_G = ( D_\text{N} \phi' , \phi )_G
   \ , \\
   ( \phi , D_\text{N} \phi )_G \ge 0
   \ , \\
   ( \phi , D_\text{N} \phi )_G = 0 \ \Leftrightarrow \ \phi = 0
   \ .
\end{gather}
The desired operator is defined as
\begin{gather}
   \Delta = G^{1/2} D_\text{N} G^{-1/2} \ .
\end{gather}
Symmetry and strict positivity of $\Delta$ with respect to the canonical scalar
product of $\Delta$ follow from the corresponding properties of $D_\text{N}$.
Notice that
\begin{gather}
   D^\alpha = G^{1/2} D_\text{N}^\alpha G^{-1/2} \ .
\end{gather}

The \texttt{openQ*D} code uses the Fast Fourier Transform (FFT) algorithm to
contruct $\tilde{\pi}(p,\mu)$ from $\pi(x,\mu)$ and vice versa. The FFT is
implemented in the module \texttt{dft} which is an adaptation of the
corresponding module in the \texttt{NSPT-1.4} code written by Mattia Dalla Brida
and Martin L\"uscher~\cite{DallaBrida:2017tru}.